\documentclass[twocolumn]{aa}

\usepackage{graphicx}
\usepackage{txfonts}
\usepackage{color}
\usepackage{caption}
\usepackage{subcaption}

\usepackage{amsmath}

\usepackage{lscape}

\usepackage{natbib}
\bibpunct{(}{)}{;}{a}{}{,} %% natbib format for A&A and ApJ

\usepackage[breaklinks=true]{hyperref} %% to avoid \citeads line fills

\usepackage{enumitem}

\newcommand\T{\rule{0pt}{2.6ex}}       % Top strut
\newcommand\B{\rule[-1.2ex]{0pt}{0pt}} % Bottom strut

\begin{document}

   \title{Source clustering in the Hi-GAL\thanks{Hi-GAL is a key-project of the Herschel Space Observatory survey \citep[][]{Pilbratt2010} and uses the PACS \citep[][]{Poglitsch2010} and SPIRE \citep[][]{Griffin2010} cameras in parallel mode.} survey determined using a minimum spanning tree method}

   \author{M. Beuret\inst{1}
          \and
          N. Billot\inst{2}
          \and
          L. Cambrésy\inst{1}
          \and
          D.~J. Eden\inst{3}
          \and
          D. Elia\inst{4}
          \and
          S. Molinari\inst{4}
          \and
          S. Pezzuto\inst{4}
          \and
          E. Schisano\inst{4}
          }

   \institute{Observatoire astronomique de Strasbourg, Universit\'{e} de Strasbourg, CNRS, UMR 7550, 11 rue de l'Universit\'{e}, F-67000 Strasbourg, France\\
              \email{maxime.beuret@astro.unistra.fr}
         \and
             Instituto de Radio Astronom\'{i}a Milim\'{e}trica, Avenida Divina Pastora, 7, Local 20, 18012 Granada, Spain\\
         \and
             Astrophysics Research Institute, Liverpool John Moores University, Ic2 Liverpool Science Park, 146 Brownlow Hill Liverpool L3 5RF, UK\\
         \and
             INAF – Istituto di Astrofisica e Planetologia Spaziali, via Fosso del Cavaliere 100, I-00133 Roma, Italy\\
             }

   \date{}

% \abstract{}{}{}{}{} 
% 5 {} token are mandatory
 
  \abstract
  % context heading (optional)
  % {} leave it empty if necessary  
   {}
  % aims heading (mandatory)
   {The aims are to investigate the clustering of the far-infrared sources from the Herschel infrared Galactic Plane Survey (Hi-GAL) in the Galactic longitude range of -71 to 67 deg. These clumps, and their spatial distribution, are an imprint of the original conditions within a molecular cloud. This will produce a catalogue of over-densities.}
  % methods heading (mandatory)
   {The minimum spanning tree (MST) method was used to identify the over-densities in two dimensions. The catalogue was further refined by folding in heliocentric distances, resulting in more reliable over-densities, which are cluster candidates.}
  % results heading (mandatory)
   {We found 1,633 over-densities with more than ten members. Of these, 496 are defined as cluster candidates because of the reliability of the distances, with a further 1,137 potential cluster candidates. The spatial distributions of the cluster candidates are different in the first and fourth quadrants, with all clusters following the spiral structure of the Milky Way. The cluster candidates are fractal. The clump mass functions of the clustered and isolated are statistically indistinguishable from each other and are consistent with Kroupa's initial mass
function.}
  % conclusions heading (optional), leave it empty if necessary 
   {}

%\textbf{\color{red} catalogs (rejected and reliable source catalogs) of young stellar clusters will be provided with the paper. They will be archived electronically at the CDS.}

   \keywords{Stars:formation -- Stars:mass function -- Galaxy:structure -- Infrared:Stars -- Catalogs}

   \maketitle
%
%________________________________________________________________

   %*****************************************************************************
   %********************                                     ********************
   %********************            Introduction             ********************
   %********************                                     ********************
   %*****************************************************************************

\section{Introduction}

In recent years, the development of Galactic Plane surveys has allowed for the statistical properties of the star-formation process to be examined because these surveys encompass many different Galactic environments. These environments cover all length scales, from kpc, which allows studying interactions in the spiral arms, to pc scales, which enable us to study individual star-forming regions. Studies have started to show that, on large scales, there are no major variations in the star-formation efficiency caused by the spiral arms \citep[][]{Moore2012, Eden2013, Eden2015} but that the sub-10-pc scales may be the most important, with the most significant variations found on these smaller scales \citep[][]{Eden2012, Vutisalchavakul2014}. These results point towards the smaller scales as the most important, and therefore local triggering may be vital in the star-formation process \citep[e.g.][]{Deharveng2005, Thompson2012, Kendrew2012}.

\citet{Billot2011} used the Herschel infrared Galactic Plane Survey \citep[Hi-GAL; ][]{Molinari2010a, Molinari2016} science demonstration phase (SDP) fields to study the clustering of star-forming clumps, within two 2x2 square deg fields. In this paper, we extend this to cover $\sim$ 1/3 of the Galactic Plane, in the Galactic longitude range of -71 deg to 67 deg, spanning the first and fourth quadrants \citep[][]{Molinari2016}.

%("clump" is preferred to "core" because most of sources are not individual sources)

We use the minimum spanning tree (MST) method to investigate the clustering of star-forming clumps traced at far-IR-submm wavelengths. Single-band catalogues at 70, 160, 250, 350, and
500 have been merged together \citep{Elia2016} to produce a band-merged catalogue that is the starting point of our cluster analysis. A byproduct of identifying the clumps that fall into these clusters is that field clumps will also be identified, allowing for a comparison between these two environments. Measurements of the stellar initial mass function (IMF) in the Galaxy and in extragalactic structures find that the IMF is invariant, with no significant differences found \citep[][]{Bastian2010}. Observers have found \citep[e.g. ][]{Beltran2006, Simpson2008} that the clump mass function (CMF) shape matches the IMF either as slope or turnover mass once a constant mass offset is applied. With this observational evidence, any changes detected in the CMFs of different environments may therefore indicate a change in the IMF.

\begin{figure}
\centering
\begin{subfigure}[b]{9cm}
   \resizebox{\hsize}{!}{\includegraphics{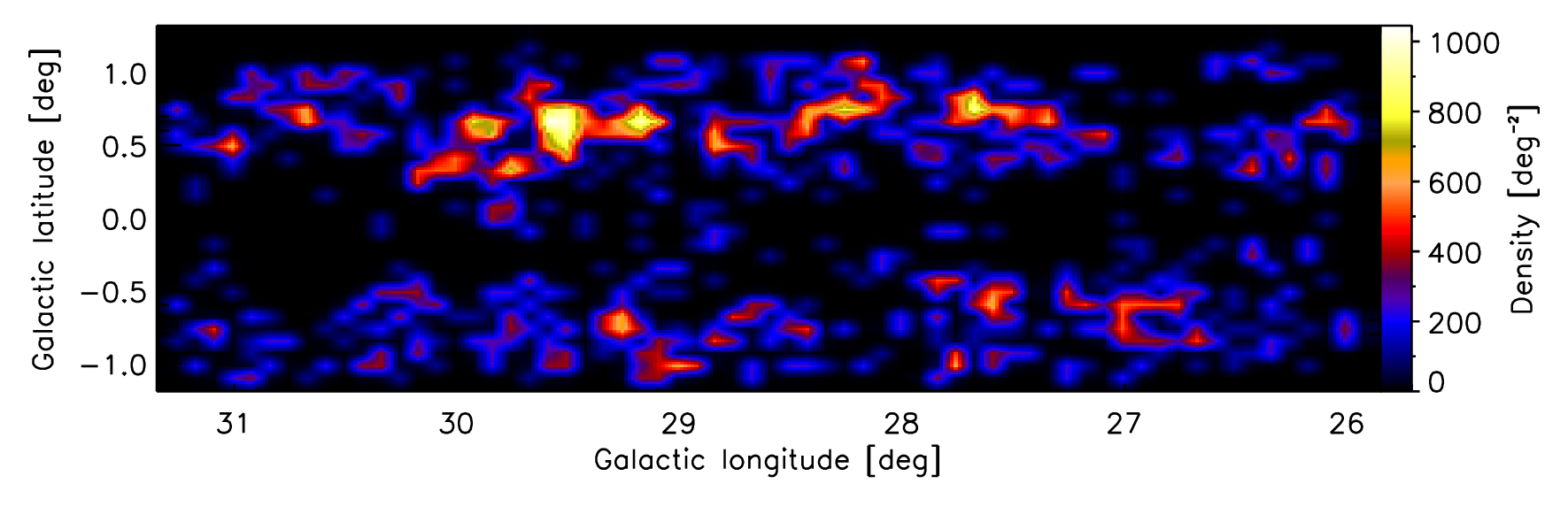}}
   \subcaption{Unbound clumps.}
   \label{densityunbound}
\end{subfigure}%
\quad
\begin{subfigure}[b]{9cm}
   \resizebox{\hsize}{!}{\includegraphics{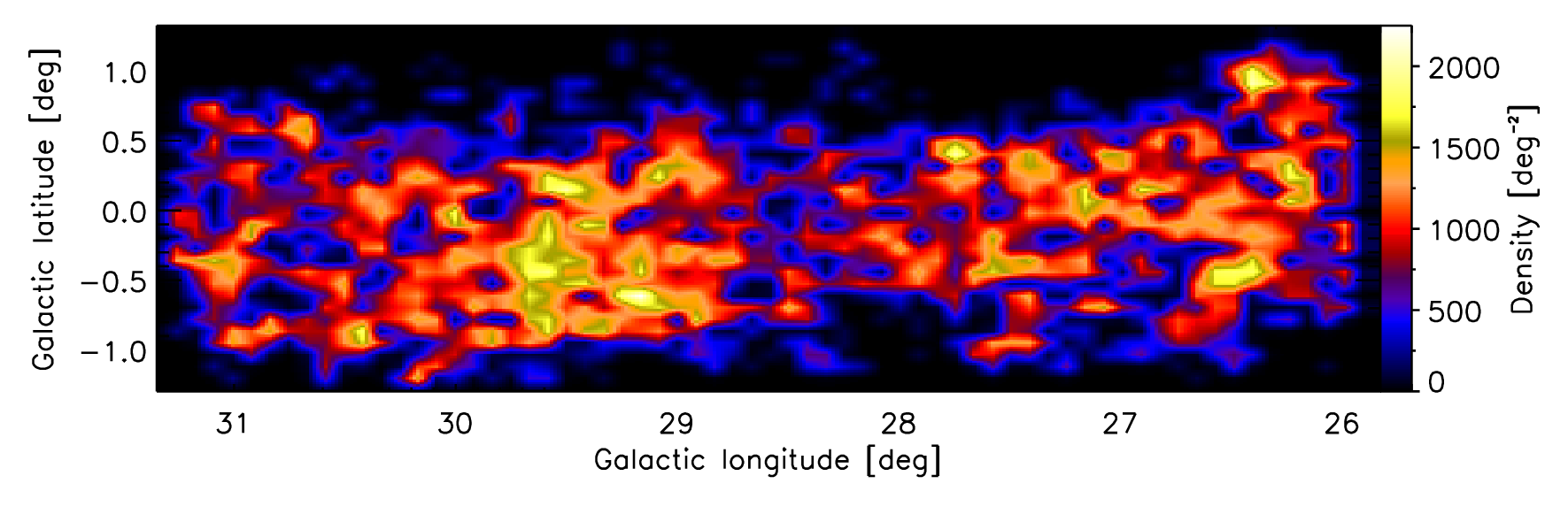}}
   \subcaption{Pre-stellar clumps.}
   \label{densityprestellar}
\end{subfigure}
\quad
\begin{subfigure}[b]{9cm}
   \resizebox{\hsize}{!}{\includegraphics{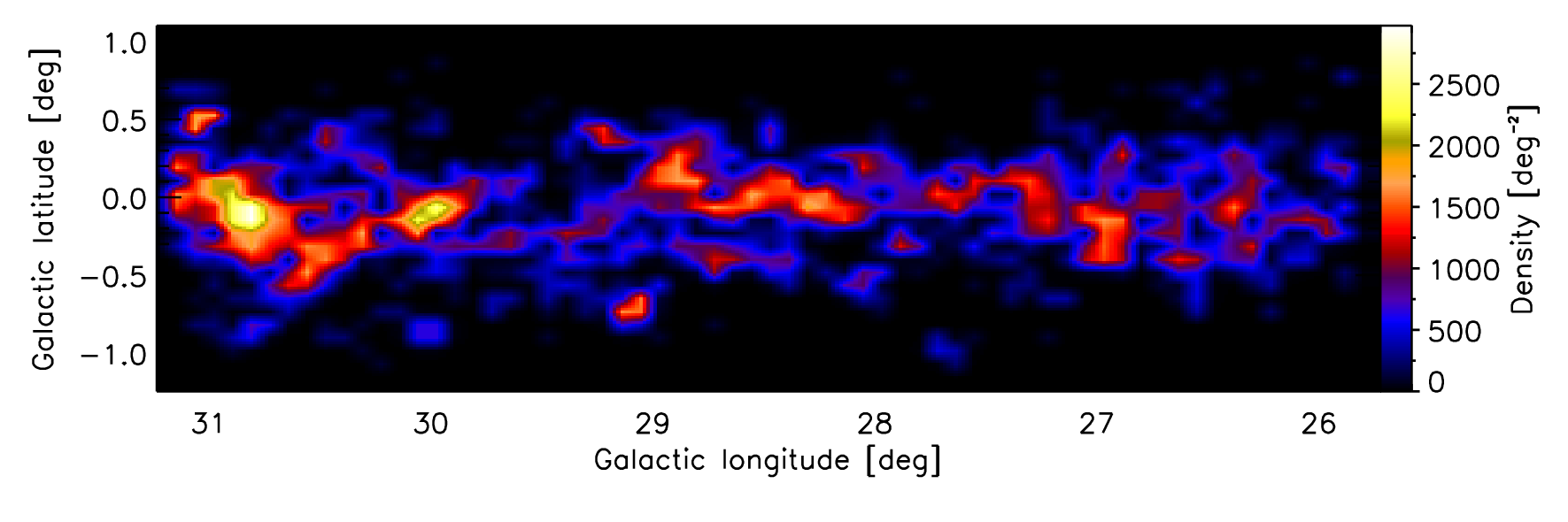}}
   \subcaption{Proto-stellar clumps.}
   \label{densityprotostellar}
\end{subfigure}
\caption{Source density maps of a sub-sample of the three different types of clumps defined in the Hi-GAL product catalogue, located in the longitude range 26 $\leq$ $\emph{l}$ $\leq$ 31~deg. The same distributions are observed for the whole sample.}
\label{densityclumps}
\end{figure}

The band-merged Hi-GAL product catalogue \citep{Elia2016} is built as in \citet{Elia2013} and provides spectral energy distribution (SED) fit parameters to the individual clumps. The average angular size of the clumps is 25\arcsec at 250~$\mu$m. Using the heliocentric distances provided in the Hi-GAL product catalogue (described in Sect. 3.2.) and the SED fit parameters, the authors of the catalogue are able to provide linear diameters and masses of the clumps. The catalogue explores a wide range of linear diameters and masses, from sub-parsec ($\leq 0.1$~pc) to parsec scale ($1-5$~pc) with masses from $1~M_\odot$ to $10^5~M_\odot$. These wide ranges mean that we probably mix several types of objects, from single star-forming cores to clumps containing multiple cores, even to entire clouds, depending on the distance of object. Most of these sources, however, fulfil the definition of clump, according to the definition of \citet{Bergin2007}. Dust temperatures for these clumps have been estimated through a grey-body fit, and searched in the range T=5-40~K. Most of them are found between 10 and 20~K. The appearance of the SED and the parameters obtained through the grey-body fit allow for classifying the evolutionary stage of these objects. Three stages are identified: starless unbound and bound (pre-stellar) objects, and proto-stellar objects. The pre- and proto-stellar stages are distinguished from each other by the presence of a 70~$\mu$m source in a proto-stellar clump \citep[e.g. ][]{Dunham2008,Ragan2012,Veneziani2013}. The bound versus unbound identification is obtained by using the mass-radius relation, well known as Larson's third law, originally formulated as $M(r) > 460M_\odot (r/pc)^{1.9}$, with $r$ the radius of the source \citep{Larson1981}. Beyond 4-5~kpc two effects could lead to misclassifying the pre- and proto-stellar stages. First, different sensitivities of PACS and SPIRE could lead to missing a possible 70~$\mu$m counterpart of a source detected with SPIRE. Second, at large heliocentric distances, two or more pre- and proto-stellar sources could be detected as a single object as a result of lacking resolution, globally and simply labelled as proto-stellar. The first effect was partially mitigated by searching for a possible 70~$\mu$m counterpart that was not originally listed in the single-band catalogue through performing additional source detection at this band using a threshold less demanding than the initial one. \citet{Elia2016} provide statistics and a discussion about the ratio between pre- and proto-stellar clumps. The distribution of the three evolutionary stages is shown in a portion of the Galactic Plane in Fig.~\ref{densityclumps}. Each panel represents an evolutionary stage in the longitude range 26 $\leq$ $\emph{l}$ $\leq$ 31~deg. The pre-stellar clumps are more extended in Galactic latitude than the proto-stellar clumps. The unbound clump distribution is hard to characterise because it is obscured by the proto-stellar clumps in the mid-plane, therefore we only consider clustered over-densities composed of pre- and proto-stellar clumps. These distributions are observed across the entire longitude range of this study.

%calculating the Bonnor-Ebert mass, $M_\text{BE} \approx 2.4 R_\text{BE} a^2/ G$ \footnote{$a=\sqrt{k_\text{b} T/\mu}$ where $k_\text{b}$ is the Boltzmann constant, $T$ the clump temperature and $\mu$ the mean molecular weight. $R_\text{BE}$ is the Bonner-Ebert radius which can be approximated by the radius of the clump. $G$ is the gravitational constant.} \citep[][]{Giannini2012}.

Additional evolutionary markers, such as \ion{H}{II} regions, a marker of high-mass star formation \citep[][]{Urquhart2013} and infrared dark clouds (IRDC), can be associated with the clustered clumps, giving further insight into the environmental conditions.

This paper is arranged as follows. Section 2 introduces the MST method, allowing the clustered clumps to be identified, with Sect. 3 describing the process of applying the method to the Hi-GAL data. In Sect. 4 we present the results of the study, with Sect. 5 comprising the discussion, whilst Sect. 6 presents the catalogue of over-densities. Finally, Sect. 7 contains the summary and conclusions.

   %*****************************************************************************
   %********************                                     ********************
   %********************    Minimum Spanning Tree method     ********************
   %********************                                     ********************
   %*****************************************************************************

\section{Minimum spanning tree method}

Clumps in these wavelengths are mostly associated with filaments \citep[e.g.][]{Molinari2010b}. For this reason, we decided to use an MST method, which is well suited to find sources along filamentary structures. This method was first described in \citet{Boruvka1926a, Boruvka1926b}, with an English translation provided in \citet{Nesetril2010}. Since this time, several algorithms have been described, with \citet{Graham1985} providing a detailed historical evolution of the MST algorithm. For this study we used Prim's algorithm \citep{Prim1957}. More details about Prim's algorithm can be found in \citet{Schmeja2011}.

The MST method belong to undirected graph theory. Using Delaunay triangulation \citep[e.g.][]{Shamos1975, Toussaint1980}, the method connects all points, called vertices, with branches, or edges, without creating closed loops whilst minimising the total length of the branches. As all lengths are different, the solution of the MST analysis is unique. In this case, the vertices correspond to the clumps, and the edges, $\Lambda$, correspond to the angular distance separating the clumps regarding the solution found with the MST method. This algorithm was originally used in an astrophysical context for galaxy clusters. More recently, it was used in stellar clusters, for instance in \citet{Koenig2008}, \citet{Gutermuth2009}, \citet{Billot2011} and \citet{Saral2015}.  This method was also recently used to observe the mass segregation into star clusters in \citet{Allison2009}, \citet{Maschberger2011}, and \citet{Parker2012}.

For this study we followed the recommendations of \citet{Koenig2008} and \citet{Gutermuth2009} to analyse the trees that are determined. By producing the cumulative distribution of the branch lengths, an estimate of the branch cut-off length, $\Lambda_\text{cut}$, can be found. Two segments are fitted to each extreme of the cumulative distribution with the first to the small branches, whereas the other one will fit the large branches. The intersection of these two segments provides a branch cut-off length $\Lambda_\text{cut}$. Differently from other methods where the cut-off threshold is chosen manually, this method allows an automatic setting of the cut-off.

\begin{figure}
\resizebox{\hsize}{!}{\includegraphics{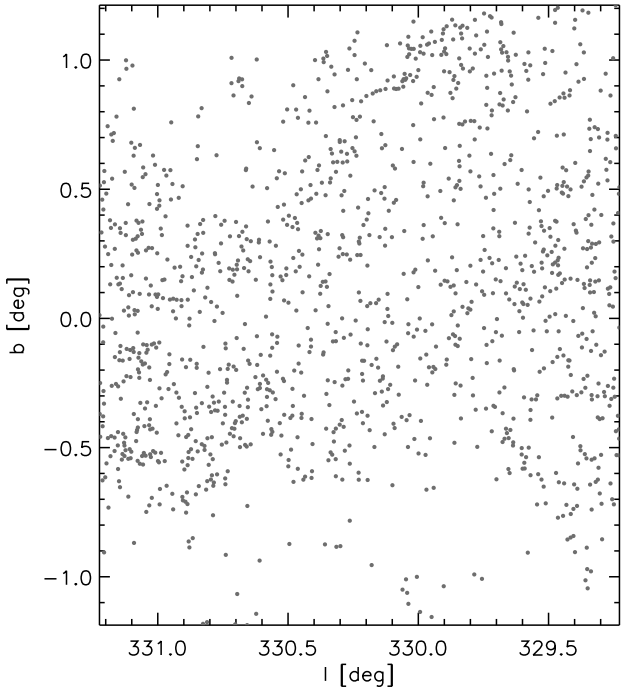}}
\caption{Distribution of pre-stellar and proto-stellar clumps in a 2 $\times$ 2 square deg field centred on $\emph{l}$ = 330~deg.}
\label{Distrib}
\end{figure}

\begin{figure}
\resizebox{\hsize}{!}{\includegraphics{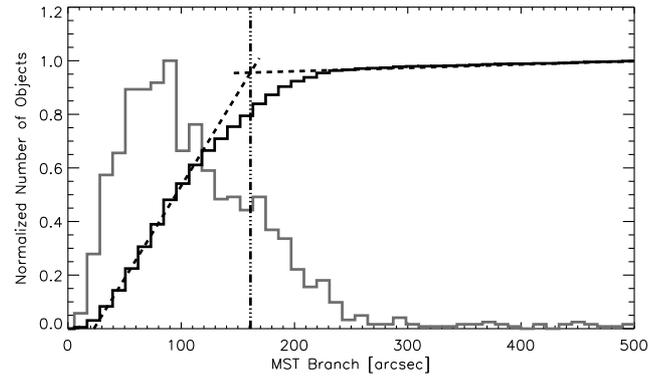}}
\caption{Distribution of the branch lengths found in the field displayed in Fig.~\ref{Distrib}. The grey line corresponds to the branch length histogram, whilst the black line is the cumulative distribution. The two dashed black lines are the fitted segments to the cumulative distribution. The dashed-dot line shows the cut-off branch length found at the intersection of these fits.}
\label{Branchdistrib}
\end{figure}

\begin{figure}
\centering
\resizebox{\hsize}{!}{\includegraphics{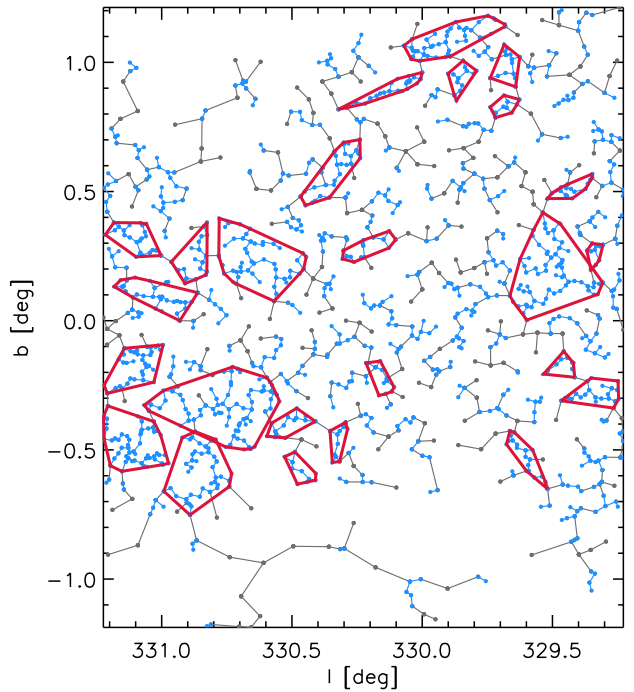}}
\caption{Result of the MST method applied on the field in Fig.~\ref{Distrib}. Grey points and lines correspond to $\Lambda>\Lambda_\text{cut}$ , whereas blue points and lines correspond to $\Lambda<\Lambda_\text{cut}$. Red lines correspond to convex hulls surrounding groups with $N \ge 10$.}
\label{msthigal}
\end{figure}

Figure~\ref{Distrib} shows the distribution of Hi-GAL proto-stellar and pre-stellar clumps for a 2 $\times$ 2~$\text{deg}^2$ region of the sky centred on l=330~deg. From this sub-sample, we can determine the branch cut-off length by using the branch distribution as shown in Fig.~\ref{Branchdistrib}. Figure 3 shows the cumulative distribution (in black) from the histogram of branches (in grey) of this sub-sample. The intersection of the two linear fits corresponds to a $\Lambda_\text{cut}$ of 160\arcsec for this sub-sample. Only branches smaller than $\Lambda_\text{cut}$ are considered. The result for this field is presented in Fig.~\ref{msthigal}. Over-densities are represented by their convex hulls. Some hulls seem to overlap others, but this appears to be a problem of representation of the correct shape for the over-densities. Some points that should belong to a convex hull are ignored and a straight line is drawn instead of several segments. A clump belongs to only one over-density and always respects the criterion of the cut-off branch length.
Following the recommendation of \citet{Gutermuth2009}, we applied a threshold of the minimum number of clumps required to form an over-density, using a threshold of $N=10$ clumps. Most of the \textit{\textup{visual}} over-densities in Fig.~\ref{Distrib} can clearly be found by the MST method.

   %*****************************************************************************
   %********************                                     ********************
   %********************      Catalog of over densities      ********************
   %********************                                     ********************
   %*****************************************************************************

\section{Cluster catalogue definition}

In this section we describe how we applied the method defined in Sect.2 to the Hi-GAL catalogue. The whole catalogue is analysed in pieces within a rectangular window. The choice of the window affects the results in terms of cut-off threshold and number of clusters. The heliocentric distance estimates help us to distinguish probable clusters from the list of over-densities.\\

\subsection{Effect of window size}

\begin{figure}
\resizebox{\hsize}{!}{\includegraphics{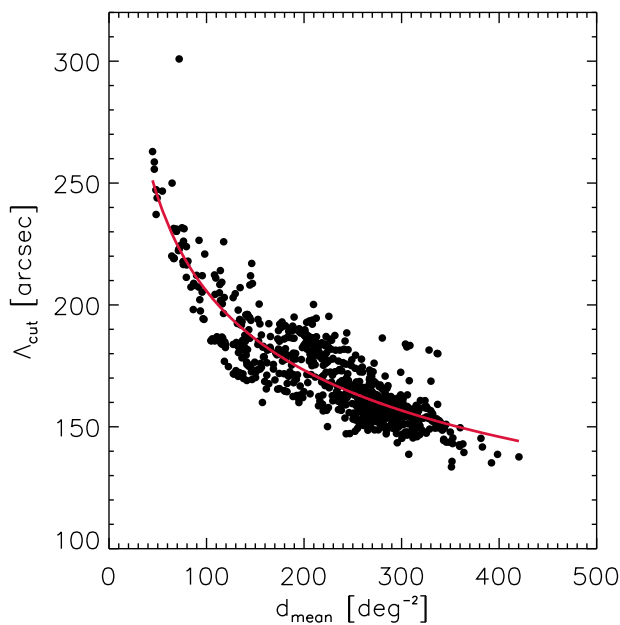}}
\caption{Relation between $\Lambda_\text{cut}$ and the mean density of the field considered. The red line corresponds to the correlation function $\Lambda_\text{cut} \propto d_\text{mean}^{-0.24}$.}
\label{lcutdens}
\end{figure}

The determination of $\Lambda_\text{cut}$ is associated with the mean density of the sample. Higher mean densities result in lower values of $\Lambda_\text{cut}$. Figure~\ref{lcutdens} shows the $\Lambda_\text{cut}$ distribution along the mean density. Each point corresponds to a solution of the MST for different fields and different ranges of longitude. The correlation between $\Lambda_\text{cut}$ and the mean source density is well represented by a power law, $\Lambda_\text{cut} \propto d_\text{mean}^{-\beta}$ with a slope $\beta \approx 0.24 \pm 0.01$. The spread in the distribution highlights a strength of the MST method. The MST method can find over-densities even against a very busy background distribution, therefore the detection of over-densities depends on the density contrast between the over-densities and the background, as was reported by \citet{Schmeja2011}.

The source density varies across the Galactic Plane, mainly as a result of Galactic structure features such as the Galactic Centre and the spiral arms, which will increase the number of sources. These variations are not seen dramatically in the latitude direction. The latitude range (2 deg) is smaller than the longitude range, with no large spatial variations due to Galactic structure.

This density variation along the longitude forced us to split the whole catalogue into rectangular windows. By not analysing the whole Galactic Plane together, fields need to be overlapped to avoid rejecting clumps that lie at the edges of fields. To avoid this, a new window was created by shifting a field in Galactic longitude by half. For each window, a branch histogram was built that determined a $\Lambda_\text{cut}$ value. However, this potentially resulted in two values of $\Lambda_\text{cut}$ and two different shapes for the over-density. As a result, the window in which the over-density was completely contained dictated the choice.

\begin{figure*}
\centering
\includegraphics[width=\textwidth,clip]{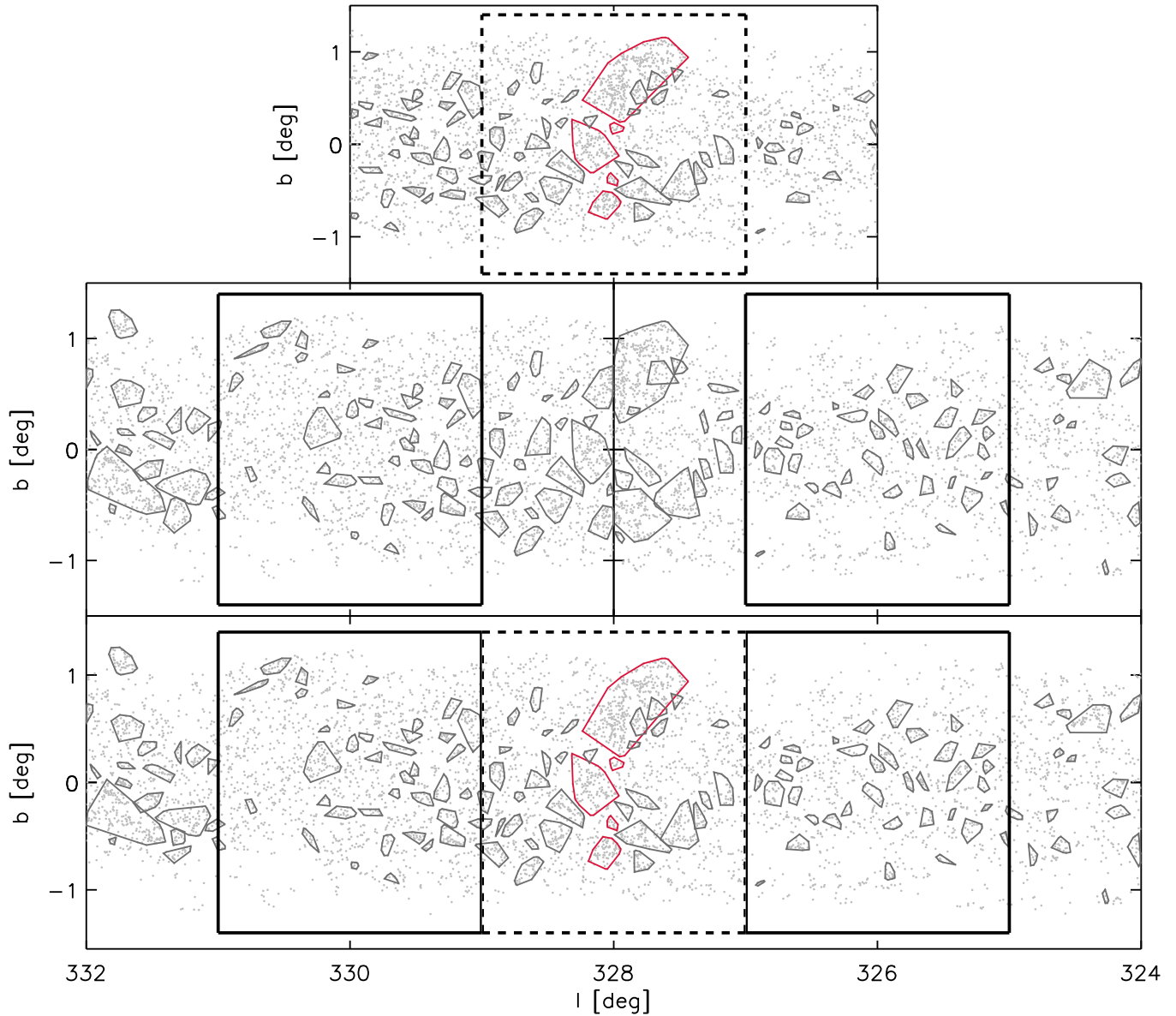}
\caption{Example of the reconstructing of the over-density
catalogue found by the MST method at the split between windows. This example field is 10 deg large. Grey points are the pre-stellar and proto-stellar clumps. The grey shapes correspond to the convex hulls of the over-densities with at least ten members. The red shapes are the convex hulls of the over-densities split over the two original windows. The black dashed and solid lines mark the separation limits for each window. The top window is the overlap window. The middle windows are the two original windows of this field. The bottom window is the result of the overlapping method.}
\label{overlap}
\end{figure*}

Figure~\ref{overlap} illustrates the process of rebuilding the whole catalogue. The top panel contains the overlap window, whilst the middle panel contains the two original windows. The lower panel houses the final catalogue for this Galactic longitude range. The grey shapes correspond to the convex hulls of the over-densities found in each window with at least ten members. The red shapes correspond to the convex hulls for the over-densities split into two by the original windows. The boxes correspond to the limits of the areas considered in the catalogue-building process.  

As a result of the overlap, the choice of window size may have a strong effect on the result. The window size also needs to correspond to the spatial distribution of the data. To do this, a range of 2 to 30 deg was investigated, stepping along the data by a step of one window in size. The minimum size of 2 deg was chosen as this is the size of the Hi-GAL observation fields. This size also allows for a varied distribution in branch sizes, and is larger than the typical size of an over-density as displayed in Fig.~\ref{Branchdistrib}. Therefore, the window size of 2 deg is large enough to conserve most branch lengths. The maximum value of 30 deg was chosen to be larger than tangents of spiral arms and the Galactic Centre.

\begin{figure}
\centering
\resizebox{\hsize}{!}{\includegraphics{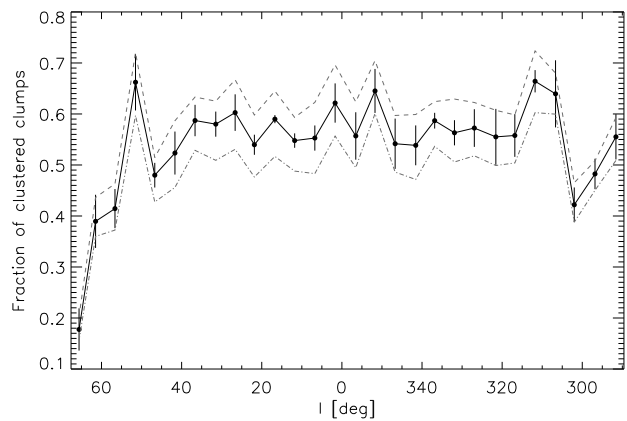}}
\caption{Spatial distribution of the fraction of clustered clumps as a function of Galactic longitude. The black line corresponds to the mean fraction of clustered clumps of each master catalogue (one for each window size) for each 5 deg bin. The black points correspond to this mean and are located at the average longitude of each bin. The 1$\sigma$ error bars correspond to the standard deviation of the fraction of clustered clumps for each bin. The grey dashed and dot-dashed lines correspond to a value of $\pm$ 5\% of $\Lambda_\text{cut}$.}
\label{effectwindow}
\end{figure}

Furthermore, we evaluated the non-random clustering of sources, that is, whether the over-densities are real. Using the approach of \citet{Campana2008}, the total length of an MST analysis, $\Lambda$, is proportional to $\sqrt{AN_{tot}}$ , where $A$ is the window area and $N_{tot}$ is the total number of clumps inside this window, and the mean length is proportional to $\sqrt{A/N_{tot}}$. For a random-field MST analysis, Campana and collaborators found coefficients of $\approx 0.65$ for both quantities from Monte Carlo simulations. We find mean coefficients of $\approx 0.51 \pm 0.01$ and $\approx 0.40 \pm 0.01$ for the total length and mean length, respectively, over all window sizes checked in this study. These values are significantly lower than what would be expected for a random-MST field, which means that the sources are distributed non-randomly. This result allows all window sizes to be considered.

Figure~\ref{effectwindow} shows a test of the effect of the window size as a function of Galactic longitude. By building 29 catalogues, one for each window size between 2 and 30 deg, and computing the fraction of clumps in over-densities, we plotted the mean of the 29 catalogues of this fraction in 5-deg bins. The error bars correspond to a 1$\sigma$ error of the standard deviation in each bin, with the dashed and dot-dashed lines representing a variation of $\pm$ 5\% of the branch cut-off length, which is approximately 10\arcsec and represents a variation of 10-15\% of the amount of clustered sources. This distribution shows that there is not much variation, other than at the edges of the longitude range, therefore the window size does not cause too much variation.

\begin{figure*}
\centering
\begin{subfigure}[b]{9cm}
   \resizebox{\hsize}{!}{\includegraphics{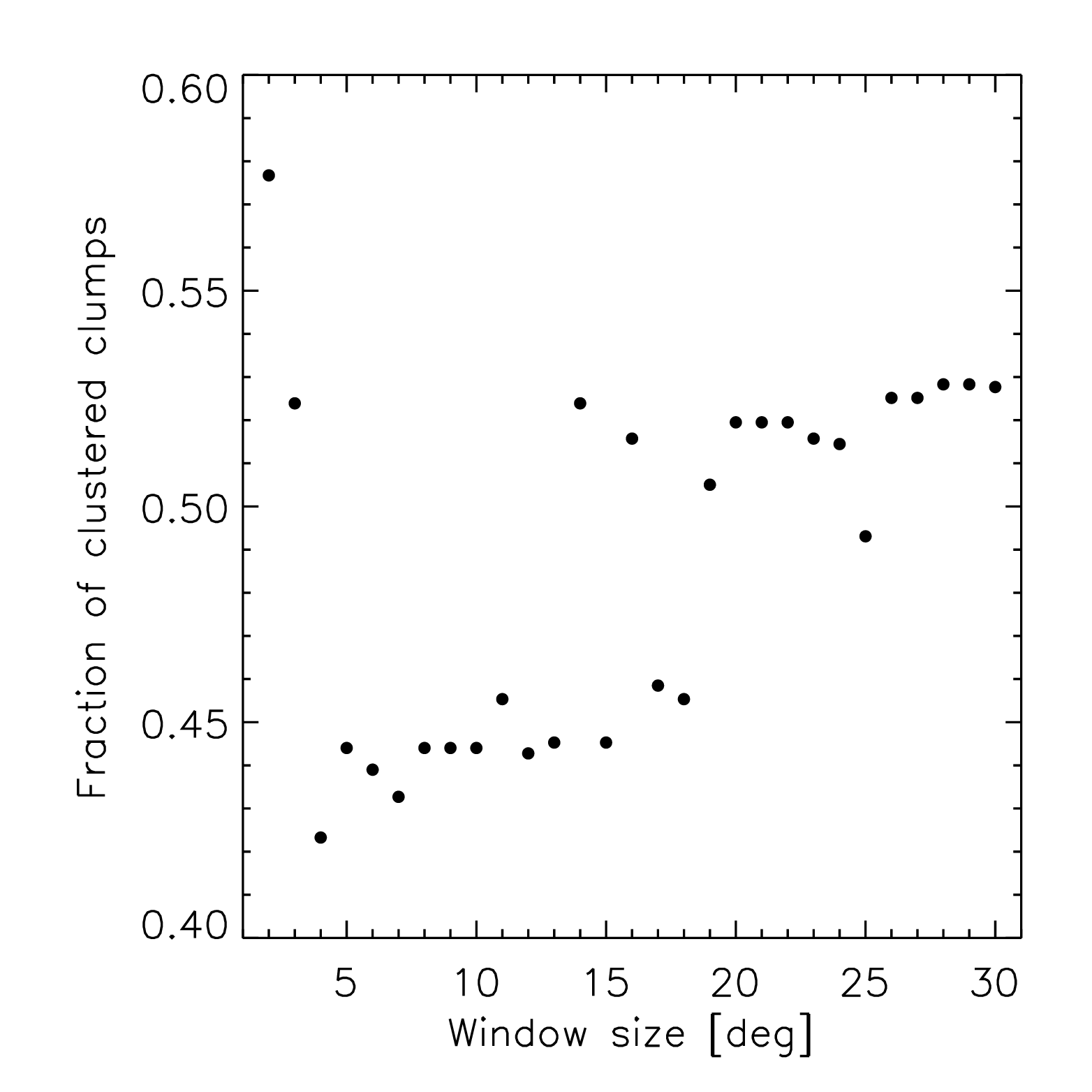}}
   \subcaption{2 $\times$ 2~deg field centred on $\emph{l}$ = 335 deg.}
   \label{sigmalowdens}
\end{subfigure}%
\quad
\centering
\begin{subfigure}[b]{9cm}
   \resizebox{\hsize}{!}{\includegraphics{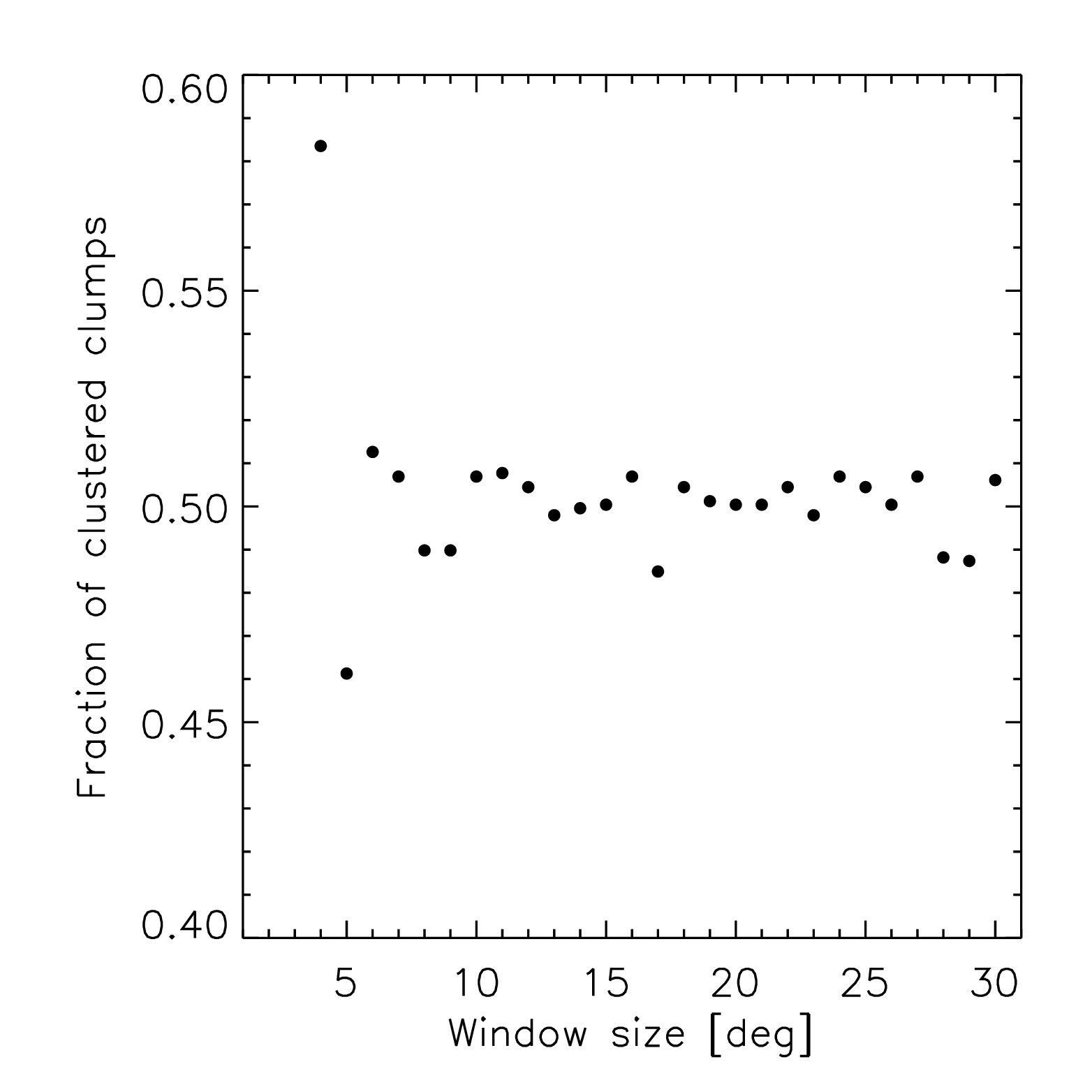}}
   \subcaption{2 $\times$ 2~deg field located at $\emph{l}$ = 20~deg.}
   \label{sigmahighdens}
\end{subfigure}%
\caption{Variation of the fraction of clustered clumps into 2 $\times$ 2 square deg fields for all window sizes between 2 and 30 deg.}
\label{fracclustclumps}
\end{figure*}

As the clustering fraction does not vary over the whole study region, we need to look at particular regions and study the effect of window size on these smaller regions. Again, only over-densities with over ten members were considered. Figures~\ref{sigmalowdens} and \ref{sigmahighdens} show two particular regions with a size of 2x2 deg, centred on 335~deg (Fig.~\ref{sigmalowdens}) and 20~deg (Fig.~\ref{sigmahighdens}). The distribution of the clumps is different in the two regions, with the field centred on 335~deg having a higher source density than at 20~deg. In both distributions, one window size does not vary the fraction of clustered clumps, but in the higher density 335~deg window, using a window larger than $\sim$ 20~deg causes a greater jump in the fraction found. This fraction of clustered clumps is influenced by the cut-off threshold and by both the number and distribution of sources that lie within the window size.

\begin{figure}
\resizebox{\hsize}{!}{\includegraphics{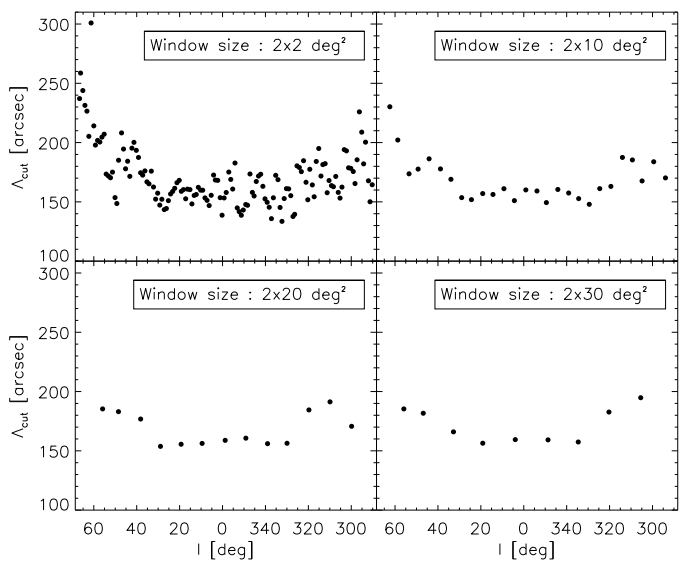}}
\caption{Changes in the $\Lambda_\text{cut}$ as a function of Galactic longitude in four different window sizes, with sizes of 2, 10, 20, and 30 deg. The Galactic longitude corresponds to the centre of each window.}
\label{lcutglon}
\end{figure}

It is also pertinent to try to remove small-scale variations, that is, those below 2 deg. Figure~\ref{lcutglon} shows the variation of $\Lambda_\text{cut}$ along the Galactic longitude for four different window sizes, 2, 10, 20, and 30 deg. A larger window produces a smoother distribution of $\Lambda_\text{cut}$, with windows larger than 20 deg smoothing out any variation in source densities caused by Galactic structure. Comparing the results from Figs.~\ref{sigmalowdens}, \ref{sigmahighdens}, and \ref{lcutglon}, the best range of window sizes appears to be 10-15 deg, with 10 deg chosen for this study. As a result, a catalogue of 1,705 over-densities with at least ten members was found, which corresponds to 45323 clustered clumps.

\subsection{Heliocentric distance estimates}

The 1,705 over-densities characterised are distributions of clumps in two dimensions. As the Galaxy is three-dimensional, we wish to remove the casual clustering induced by projection along the line of sight. This selection in the catalogue cluster candidates can be made by using heliocentric distance estimates (named HDEs in the following) from the Hi-GAL product catalogue \citep{Elia2016}. These HDEs were extracted from the \element[][12]{CO} or \element[][13]{CO} spectrum for each clump. The authors used data from the Five College Radio Astronomy Observatory (FCRAO) Galactic Ring Survey \citep[GRS, ][]{Jackson2006} and the Exeter-FCRAO Survey (Brunt et al., in prep.; Mottram et al., in prep.) for the first quadrant. The NANTEN \element[][12]{CO} data were used for the fourth quadrant. When needed, the distance ambiguity was resolved by using the extinction maps and a catalogue of sources with known distances (\ion{H}{II} regions, for example). A case study calculating kinematic distances in the Hi-GAL SDP fields can be found in \citet{Russeil2011}.
Of the 99,083 Hi-GAL sources that cover all evolutionary phases, 57\% have HDEs. Of the population of proto-stellar clumps, 64\% have distance estimates. HDEs have not been obtained for sources in the longitude range -10 to +14~deg where kinematic distances cannot be estimated. When this range is discounted, 80\% of the proto-stellar clumps have HDEs.

It is not suitable to use the distances to compute the MST analysis in three dimensions as the uncertainties on HDEs are very large \citep[$\approx$0.6 kpc and $\approx$0.9 kpc for the SDP $\emph{l}$ = 30 deg and $\emph{l}$ = 59 deg;][]{Russeil2011} as \citet{Billot2011} discuss. This would result in rejecting large parts of the catalogue.

As a result, the MST was computed and the HDEs were used as a third dimension to evaluate the compactness of each over-density. For each over-density we computed the HDE histogram and assumed all the sources to belong to the
cluster that lay within 1 kpc from its peak \citep{Russeil2011}. This range is called \textit{reliable distance}, RD, in the following.

Two metrics were used to define the compactness. The first is the ratio of the number of sources inside the RD to the total number of sources with HDEs. The second is the ratio of the number of sources with HDEs to the total number of sources. A threshold of 70\% of clumps inside the RD and 40\% of clumps with HDEs was used for the two metrics. One last measure was used: each over-density must have five clumps with HDEs. Six hundred and
seven over-densities satisfy the compactness criteria and have five clumps with HDEs, compared to the 1,705 over-densities found in two dimensions.

Only 219 (36\%) of the 607 over-densities have all clumps inside the RD. As a result, it was decided to remove the clumps that do not seem to belong to these cluster candidates. This would allow a test of the MST method, to determine whether it is able
to reproduce the same cluster candidates.

A total of 1,531 clumps were removed, with the MST analysis recomputed with the remaining 80\,128 clumps. A total of 1,633 over-densities were found, 449 of them having no clumps with HDEs. Of the remaining 1,184 over-densities, 496 satisfy the thresholds of  70\% of clumps inside the RD and 40\% of clumps with HDEs. Forty of the remaining 1,184 clumps have clumps outside of the RD. Of the final 496 over-densities, 364 have between 10 and 20 clumps ($\sim$ 70\%), with 4 over-densities having at least 70 members. The final 496 candidates are composed of 9,608 clumps, approximately 12\% of the total 80,128 clumps, with 8,248 ($\sim$ 86\%) having HDEs.

The new MST analysis can still be considered a non-random distribution as a value of 0.36 $\pm$ 0.01 was found for the mean MST length and $\sqrt{A/N_{tot}}$ , which is still below the upper limit of 0.65 found by \citet{Campana2008}.

   %*****************************************************************************
   %********************                                     ********************
   %********************     Description of the catalogs     ********************
   %********************                                     ********************
   %*****************************************************************************

\section{Description of the catalogues}

As a result of the work in Sect. 3.2, two catalogues of over-densities, or cluster candidates, were created. The first is of the 496 cluster candidates, with part of this catalogue displayed in Table~\ref{cc18}. The sources in this catalogue are the most reliable detections and are used from here on to derive properties for the analyses in Sects. 5 and 6. The other is the \textup{\textup{\textit{\textup{potential cluster catalogue}}}}, consisting of the 1,137 clusters rejected in the previous section. A part of this catalogue is displayed in Table~\ref{potclust}.

\begin{longtab}
\begin{longtable}{cccccccc}
\caption{\label{cc18} Partial list of cluster candidates. Columns [1-8].}\\
\hline
 & & & & & & & \\
\verb![1]!\tablefootmark{a} & \verb![2]!\tablefootmark{b} & \verb![3]!\tablefootmark{b} & \verb![4]!\tablefootmark{c} & \verb![5]!\tablefootmark{c} & \verb![6]!\tablefootmark{d} & \verb![7]!\tablefootmark{e} & \verb![8]!\tablefootmark{f} \\
 & & & & & & & \\
Cluster candidates name \, & \, l$_{center}$ \, & \, b$_{center}$ \, & \, a-axis \, & \, b-axis \, & \, PA \, & \,  $D_\text{peak}$ \, & \,  $\bar{D}$\\
 & (deg) & (deg) & (arcmin) & (arcmin) & (deg) & (kpc) & (kpc) \\
 & & & & & & & \\
\hline\hline
 & & & & & & & \\
\endfirsthead
\caption{continued. Columns [9-16].}\\
\hline
 & & & & & & & \\
\verb![9]!\tablefootmark{g} & \verb![10]!\tablefootmark{h} & \verb![11]!\tablefootmark{i} & \verb![12]!\tablefootmark{i} & \verb![13]!\tablefootmark{j} & \verb![14]!\tablefootmark{k} & \verb![15]!\tablefootmark{l} & \verb![16]!\tablefootmark{m} \\
 & & & & & & & \\
R & Z & a-axis & b-axis & Density & $Nb_\text{clumps}$ & $Nb_\text{proto}$/$Nb_\text{pre}$ & \ion{H}{II} \\
 (kpc) & (pc) & (pc) & (pc) & Clumps.pc$^{-2}$ &  &  & regions \\
 & & & & & & & \\
\hline\hline
 & & & & & & & \\
\endhead
\hline
\endfoot
\hline
G289.85-0.764  &  289.87  &  -0.76  &  7.36  &  2.85  &  16.4  &  8.05  &  8.05 \\
G289.57-0.656  &  289.58  &  -0.65  &  3.56  &  2.28  &  43.5  &  7.45  &  7.45 \\
G290.52-0.771  &  290.59  &  -0.78  &  9.14  &  7.54  &  162.  &  7.55  &  7.55 \\
G291.84-0.684  &  291.84  &  -0.68  &  4.21  &  4.21  &  0.00  &  9.05  &  9.05 \\
G292.27-0.015  &  292.27  &  -0.01  &  4.72  &  1.83  &  50.0  &  5.85  &  5.85 \\
G292.39-0.437  &  292.39  &  -0.43  &  4.83  &  3.76  &  4.96  &  5.45  &  5.45 \\
G290.30-0.014  &  290.30  &  -0.01  &  6.16  &  6.16  &  0.00  &  6.05  &  6.05 \\
G290.56-1.412  &  290.59  &  -1.42  &  6.23  &  2.68  &  23.5  &  7.35  &  7.35 \\
G290.77-1.371  &  290.79  &  -1.38  &  4.28  &  3.75  &  117.  &  7.45  &  7.45 \\
G290.46-0.338  &  290.45  &  -0.31  &  6.51  &  5.54  &  86.3  &  8.25  &  8.25 \\
G293.89-0.824  &  293.89  &  -0.82  &  11.8  &  7.42  &  47.2  &  10.3  &  10.3 \\
G293.60-1.616  &  293.58  &  -1.52  &  9.93  &  6.17  &  56.9  &  2.35  &  2.35 \\
G294.61-1.201  &  294.60  &  -1.20  &  4.89  &  3.07  &  178.  &  9.55  &  9.55 \\
G295.14-0.577  &  295.12  &  -0.55  &  4.87  &  3.20  &  50.2  &  10.8  &  10.8 \\
G295.15-1.293  &  295.16  &  -1.27  &  6.53  &  3.11  &  102.  &  2.25  &  2.25 \\
G295.14-1.589  &  295.12  &  -1.56  &  9.06  &  7.75  &  153.  &  2.25  &  2.25 \\
G294.91-1.667  &  294.91  &  -1.66  &  4.72  &  4.72  &  0.00  &  2.25  &  2.25 \\
G294.23-0.477  &  294.23  &  -0.47  &  6.09  &  6.09  &  0.00  &  7.25  &  7.25 \\
G294.08-1.614  &  294.08  &  -1.61  &  6.53  &  1.82  &  146.  &  2.25  &  2.25 \\
G294.50-1.621  &  294.49  &  -1.64  &  3.89  &  2.22  &  87.8  &  2.25  &  2.25 \\
G289.48+0.112  &  289.48  &  0.11  &  3.70  &  3.70  &  0.00  &  8.15  &  8.15 \\
G294.97+0.103  &  294.97  &  0.10  &  3.66  &  3.66  &  0.00  &  9.45  &  9.45 \\
G295.07+0.538  &  295.09  &  0.53  &  3.56  &  2.83  &  0.00  &  7.55  &  7.55 \\
G296.79-1.120  &  296.74  &  -1.10  &  18.0  &  10.3  &  62.1  &  9.95  &  9.95 \\
G296.85-1.457  &  296.89  &  -1.44  &  8.18  &  4.36  &  16.5  &  9.85  &  9.85 \\
G297.14-1.357  &  297.14  &  -1.36  &  8.53  &  4.09  &  6.62  &  9.75  &  9.75 \\
G297.50-0.822  &  297.49  &  -0.79  &  13.9  &  7.14  &  53.0  &  10.7  &  10.7 \\
G297.32-0.269  &  297.28  &  -0.27  &  5.88  &  3.00  &  170.  &  9.35  &  9.35 \\
G297.47-0.008  &  297.46  &  0.01  &  4.64  &  3.84  &  0.00  &  9.45  &  9.45 \\
G298.18-0.321  &  298.17  &  -0.30  &  11.1  &  6.34  &  80.7  &  11.3  &  11.3 \\
G298.74-0.019  &  298.74  &  -0.01  &  7.60  &  5.03  &  129.  &  10.2  &  10.2 \\
G298.85+0.170  &  298.85  &  0.17  &  5.55  &  5.55  &  0.00  &  10.4  &  10.4 \\
G298.38-0.069  &  298.38  &  -0.06  &  6.39  &  6.39  &  0.00  &  11.0  &  11.0 \\
G298.15-0.617  &  298.15  &  -0.61  &  7.17  &  7.17  &  0.00  &  11.9  &  11.9 \\
G299.52+0.119  &  299.52  &  0.11  &  3.37  &  3.37  &  0.00  &  7.75  &  7.75 \\
G298.91+0.461  &  298.90  &  0.46  &  6.17  &  4.21  &  82.4  &  2.55  &  2.55 \\
G300.34-0.281  &  300.34  &  -0.28  &  5.58  &  5.58  &  0.00  &  11.8  &  11.8 \\
G300.39+0.303  &  300.38  &  0.31  &  9.67  &  6.20  &  84.3  &  7.15  &  7.15 \\
G301.55-0.259  &  301.55  &  -0.25  &  6.56  &  4.40  &  1.43  &  6.75  &  6.75 \\
G302.44+0.025  &  302.42  &  0.03  &  11.9  &  7.57  &  40.1  &  5.95  &  5.95 \\
G302.72+0.190  &  302.67  &  0.13  &  9.27  &  6.52  &  49.3  &  5.55  &  5.55 \\
G302.74-0.078  &  302.74  &  -0.07  &  4.84  &  4.84  &  0.00  &  5.55  &  5.55 \\
G302.19-0.856  &  302.19  &  -0.85  &  7.38  &  7.38  &  0.00  &  12.0  &  12.0 \\
G302.50-0.728  &  302.50  &  -0.72  &  4.47  &  4.47  &  0.00  &  12.0  &  12.0 \\
G302.67-0.680  &  302.70  &  -0.70  &  6.40  &  5.19  &  62.8  &  11.9  &  11.9 \\
G303.88-0.792  &  303.88  &  -0.79  &  4.21  &  1.87  &  10.4  &  12.4  &  12.4 \\
G304.52-0.076  &  304.52  &  -0.07  &  9.06  &  9.06  &  0.00  &  7.45  &  7.45 \\
G304.91+0.018  &  304.91  &  0.01  &  4.71  &  4.71  &  0.00  &  7.25  &  7.25 \\
G305.65+0.522  &  305.65  &  0.52  &  3.97  &  3.97  &  0.00  &  6.85  &  6.85 \\
G306.28-0.033  &  306.29  &  -0.03  &  4.72  &  2.58  &  1.65  &  2.25  &  2.25 \\
G305.38-0.257  &  305.38  &  -0.25  &  4.08  &  0.93  &  178.  &  2.55  &  2.55 \\
G306.62+0.008  &  306.62  &  0.00  &  3.07  &  3.07  &  0.00  &  7.75  &  7.75 \\
G304.01+0.381  &  304.01  &  0.38  &  9.99  &  5.58  &  98.0  &  5.75  &  5.75 \\
G305.50-0.707  &  305.49  &  -0.70  &  9.37  &  3.97  &  152.  &  8.35  &  8.35 \\
8.01  &  -5.5  &  17.1  &  6.80  &  0.065  &  24  &  1.00  &  G289.799-00.839$^{(1)}$ \\ % columns [9-16]
9.72  &  -4.5  &  7.71  &  5.14  &  0.088  &  11  &  0.57  &  G289.582-00.636$^{(1)}$ \\
2.46  &  -5.3  &  20.0  &  16.4  &  0.023  &  24  &  1.00  &  290.5-00.8$^{(2)}$ \\
8.08  &  -5.7  &  11.2  &  11.2  &  0.035  &  14  &  3.66  &  G291.614-00.525$^{(1)}$ \\
9.39  &  -0.0  &  8.07  &  2.85  &  0.137  &  10  &  0.25  &    \\
9.82  &  -2.3  &  7.75  &  5.95  &  0.082  &  12  &  0.71  &    \\
4.78  &  -0.0  &  10.8  &  10.8  &  0.043  &  16  &  0.23  &    \\
2.54  &  -7.2  &  13.4  &  5.67  &  0.050  &  12  &  0.20  &    \\
1.26  &  -7.3  &  9.27  &  8.13  &  0.063  &  15  &  0.25  &    \\
3.56  &  -2.5  &  15.6  &  13.3  &  0.030  &  20  &  0.42  &    \\
18.8  &  -7.6  &  35.7  &  22.3  &  0.019  &  49  &  1.45  &  G293.898-00.826$^{(1)}$ \\
10.6  &  -2.3  &  6.78  &  4.21  &  0.311  &  28  &  1.33  &  G293.613-01.291$^{(1)}$ \\
17.3  &  -8.9  &  13.5  &  8.72  &  0.042  &  16  &  0.33  &  G294.793-01.330$^{(1)}$ \\
16.5  &  -5.7  &  15.4  &  9.91  &  0.039  &  19  &  1.11  &  G295.163-00.675$^{(1)}$ \\
9.65  &  -2.1  &  4.25  &  2.05  &  0.545  &  15  &  0.15  &  G294.793-01.330$^{(1)}$ \\
9.72  &  -2.2  &  5.91  &  5.09  &  0.422  &  40  &  0.90  &  G294.793-01.330$^{(1)}$ \\
10.0  &  -2.2  &  3.10  &  3.10  &  0.461  &  14  &  0.75  &  G294.793-01.330$^{(1)}$ \\
15.6  &  -3.3  &  12.7  &  12.7  &  0.046  &  24  &  0.71  &  G294.206-00.464$^{(1)}$ \\
10.7  &  -2.2  &  4.25  &  1.09  &  1.089  &  16  &  0.33  &    \\
10.5  &  -2.2  &  2.57  &  1.55  &  0.795  &  10  &  1.50  &  G294.517-01.626$^{(1)}$ \\
10.7  &  0.91  &  8.89  &  8.89  &  0.040  &  10  &  1.00  &  G289.505+00.127$^{(1)}$ \\
15.9  &  0.97  &  10.3  &  10.3  &  0.029  &  10  &  0.42  &  G295.048+00.060$^{(1)}$ \\
13.8  &  3.82  &  7.82  &  6.38  &  0.063  &  10  &  0.00  &    \\
4.10  &  -8.8  &  52.3  &  29.9  &  0.010  &  51  &  0.96  &  G296.660-00.925$^{(1)}$ \\
2.82  &  -9.7  &  23.5  &  12.7  &  0.019  &  18  &  0.20  &    \\
1.26  &  -9.5  &  24.2  &  11.6  &  0.032  &  29  &  0.93  &  G297.089-01.343$^{(1)}$ \\
3.85  &  -7.6  &  43.4  &  22.2  &  0.018  &  55  &  1.20  &  G297.497-00.758$^{(1)}$ \\
1.35  &  -2.5  &  16.1  &  7.90  &  0.034  &  14  &  0.55  &  G297.312-00.295$^{(1)}$ \\
2.80  &  0.17  &  12.6  &  10.3  &  0.036  &  15  &  0.66  &  G297.570-00.036$^{(1)}$ \\
9.86  &  -3.4  &  36.7  &  21.1  &  0.021  &  52  &  2.46  &  G298.224-00.334$^{(1)}$ \\
13.3  &  -0.2  &  22.6  &  15.0  &  0.022  &  24  &  0.50  &  G298.529-00.251$^{(1)}$ \\
14.2  &  1.77  &  16.9  &  16.9  &  0.020  &  18  &  2.00  &  G298.756+00.059$^{(1)}$ \\
11.3  &  -0.7  &  20.5  &  20.5  &  0.013  &  18  &  1.57  &  G298.529-00.251$^{(1)}$ \\
10.1  &  -6.9  &  25.0  &  25.0  &  0.007  &  14  &  0.40  &    \\
15.0  &  0.92  &  7.56  &  7.56  &  0.061  &  11  &  0.83  &    \\
9.27  &  1.14  &  4.57  &  3.17  &  0.350  &  16  &  0.23  &  G298.924+00.473$^{(1)}$ \\
20.3  &  -3.2  &  19.1  &  19.1  &  0.021  &  25  &  1.27  &  G300.084-00.485$^{(1)}$ \\
15.6  &  2.24  &  20.0  &  12.8  &  0.040  &  33  &  0.17  &  G300.566+00.169$^{(1)}$ \\
12.4  &  -1.7  &  12.9  &  8.72  &  0.050  &  18  &  0.50  &  G301.615-00.340$^{(1)}$ \\
7.43  &  0.22  &  20.6  &  13.1  &  0.067  &  58  &  0.61  &  G302.270+00.073$^{(1)}$ \\
5.96  &  0.72  &  14.9  &  10.4  &  0.062  &  31  &  1.81  &  G302.631+00.030$^{(1)}$ \\
5.57  &  -0.4  &  7.90  &  7.90  &  0.066  &  13  &  0.18  &  G302.582-00.083$^{(1)}$ \\
12.4  &  -9.1  &  25.9  &  25.9  &  0.011  &  24  &  0.71  &    \\
9.92  &  -8.0  &  15.5  &  15.5  &  0.024  &  19  &  1.11  &  G302.503-00.762$^{(1)}$ \\
8.12  &  -7.7  &  22.2  &  17.9  &  0.019  &  24  &  2.00  &  G302.614-00.756$^{(1)}$ \\
5.93  &  -8.8  &  15.4  &  6.07  &  0.043  &  13  &  1.60  &  G303.872-00.792$^{(1)}$ \\
8.19  &  -0.5  &  19.5  &  19.5  &  0.030  &  37  &  0.54  &  G304.465-00.023$^{(1)}$ \\
10.5  &  0.13  &  10.0  &  10.0  &  0.031  &  10  &  0.11  &    \\
13.7  &  3.41  &  7.84  &  7.84  &  0.072  &  14  &  0.40  &  G305.637+00.515$^{(1)}$ \\
10.6  &  -0.0  &  3.10  &  1.73  &  0.648  &  11  &  0.37  &    \\
9.69  &  -0.6  &  3.04  &  0.88  &  1.185  &  10  &  1.50  &  G305.322-00.255$^{(1)}$ \\
16.2  &  0.06  &  7.07  &  7.07  &  0.069  &  11  &  0.22  &    \\
4.78  &  2.13  &  16.7  &  9.31  &  0.077  &  38  &  1.71  &  G303.896+00.407$^{(1)}$ \\
14.3  &  -5.4  &  22.7  &  9.56  &  0.030  &  21  &  0.50  &    \\
\hline
\multicolumn{8}{c}{
\tablefoot{
\tablefoottext{a}{Column [1] is the name of the cluster candidate. The form is G$\emph{L}$$\emph{lL}$$\emph{L}$.$\emph{l}$$\emph{l}$+$\emph{B}$$\emph{B}$.$\emph{b}$$\emph{b}$ where $\emph{L}$$\emph{lL}$$\emph{L}$.$\emph{l}$$\emph{l}$ and $\emph{B}$$\emph{B}$.$\emph{b}$$\emph{b}$ correspond to the average position in Galactic longitude and latitude, respectively, of all clustered clumps.}\\
\tablefoottext{b}{Columns [2-3] correspond to the Galactic longitude and latitude ($\emph{l}$,$\emph{b}$) of the cluster candidates determined as described in Col. [1].}\\
\tablefoottext{c}{Columns [4-5] correspond to the over-density size in arcminutes. This is computed by fitting an ellipse to each over-density. Columns [4] and [5] are the semi-major and minor axes, respectively.}\\
\tablefoottext{d}{Column [6] gives the position angle (PA) of the ellipse in degrees. This is measured clockwise from the positive direction of the longitude axis.}\\
\tablefoottext{e}{Column [7] corresponds to the heliocentric distance, $D_\text{peak}$, of the cluster candidates estimated from the peak of the HDE histogram as described in Sect. 3.2.}\\
\tablefoottext{f}{Column [8] is the mean HDE, $\bar{D}$, of the clumps within an over-density.}\\
\tablefoottext{g}{Column [9] is the Galactocentric distance computed using Eq.~\ref{equaR}.}\\
\tablefoottext{h}{Column [10] is the scale height, Z, from the Galactic Plane computed using Eq.~\ref{equaZ}.}\\
\tablefoottext{i}{Column [11-12] contains the cluster candidate sizes in pc, calculated using the semi-major and minor axes in Cols. [4] and [5] and the peak heliocentric distances.}\\
\tablefoottext{j}{Column [13] is the surface density of clumps within an over-density in units of $\text{pc}^{-2}$.}\\
\tablefoottext{k}{Column [14] contains the total number of clumps, both pre- and proto-stellar, within an over-density.}\\
\tablefoottext{l}{Column [15] corresponds to the ratio of number of pre-stellar clumps over proto-stellar clumps. A ratio of 1 corresponds to an over-density consisting of only pre-stellar clumps, whilst 0 contains no pre-stellar clumps.}\\
\tablefoottext{m}{Column [16] corresponds to the closest \ion{H}{ii} region from the centre of the candidate cluster. This column is explained in Sect. 4.3.}\\
}
}
\end{longtable}
\end{longtab}

\begin{longtab}
\begin{landscape}
\begin{longtable}{cccccccccc}
\caption{\label{potclust} Partial list of potential cluster candidates.}\\
\hline
 & & & & & & & & & \\
\verb![1]! & \verb![2]! & \verb![3]! & \verb![4]! & \verb![5]! & \verb![6]! & \verb![7]! & \verb![8]!  & \verb![9]! & \verb![10]! \\
 & & & & & & & & & \\
Overdensities name & l$_{center}$ & b$_{center}$ & a-axis & b-axis & PA & Density & $Nb_\text{clumps}$ & $Nb_\text{proto}$/$Nb_\text{pre}$ & Over-density type \\
 & (deg) & (deg) & (arcmin) & (arcmin) & (deg) & (Clumps/arcmin$^2$) & & & \\
 & & & & & & & & & \\
\hline\hline
 & & & & & & & & & \\
\endfirsthead
\caption{continued.}\\
\hline
& & & & & & & & & \\
\verb![1]!\tablefootmark{a} & \verb![2]!\tablefootmark{a} & \verb![3]!\tablefootmark{a} & \verb![4]!\tablefootmark{a} & \verb![5]!\tablefootmark{a} & \verb![6]!\tablefootmark{a} & \verb![7]!\tablefootmark{b} & \verb![8]!\tablefootmark{c}  & \verb![9]!\tablefootmark{c} & \verb![10]!\tablefootmark{d} \\
 & & & & & & & & & \\
Overdensities name & l$_{center}$ & b$_{center}$ & a-axis & b-axis & PA & Density & $Nb_\text{clumps}$ & $Nb_\text{proto}$/$Nb_\text{pre}$ & Over-density type \\
 & (deg) & (deg) & (arcmin) & (arcmin) & (deg) & (Clumps/arcmin$^2$) & & & \\
 & & & & & & & & & \\
\hline\hline
 & & & & & & & & & \\
\endhead
\hline
\endfoot
\hline
G289.78-1.314  &  289.78  &  -1.20  &  46.6  &  26.9  &  76.1  &  0.15  &  128  &  2.55  &  SD \\
G289.97-0.911  &  289.98  &  -0.93  &  12.3  &  7.88  &  106.  &  0.11  &  15  &  0.87  &  SD \\
G289.82-0.619  &  289.80  &  -0.62  &  7.20  &  3.77  &  14.0  &  0.12  &  10  &  0.25  &  SD \\
G289.88-1.157  &  289.88  &  -1.15  &  11.0  &  5.03  &  126.  &  0.24  &  11  &  2.66  &  SD \\
G290.03-1.244  &  290.03  &  -1.24  &  13.3  &  13.3  &  0.00  &  0.20  &  21  &  1.33  &  SD \\
G290.14-1.400  &  290.15  &  -1.41  &  14.6  &  8.13  &  89.9  &  0.28  &  20  &  0.05  &  SD \\
G290.25-0.823  &  290.25  &  -0.82  &  10.2  &  4.75  &  44.7  &  0.39  &  20  &  1.85  &  SD \\
G291.26-0.745  &  291.26  &  -0.74  &  31.3  &  31.3  &  0.00  &  0.19  &  122  &  1.34  &  SD \\
G291.21-0.259  &  291.19  &  -0.29  &  26.2  &  10.5  &  154.  &  0.24  &  50  &  2.33  &  SD \\
G291.76-1.037  &  291.74  &  -1.06  &  11.5  &  6.15  &  95.7  &  0.26  &  14  &  0.55  &  NDA \\
G291.97-0.972  &  291.94  &  -0.98  &  18.8  &  6.27  &  157.  &  0.16  &  22  &  2.14  &  NDA \\
G292.07-1.091  &  292.10  &  -1.11  &  18.3  &  7.33  &  177.  &  0.18  &  22  &  0.29  &  SD \\
G291.59-0.450  &  291.57  &  -0.47  &  30.1  &  22.4  &  102.  &  0.22  &  101  &  4.31  &  SCC \\
G291.79-0.412  &  291.79  &  -0.41  &  12.9  &  12.9  &  0.00  &  0.24  &  20  &  Inf  &  SD \\
G291.97-0.167  &  291.97  &  -0.16  &  18.4  &  18.4  &  0.00  &  0.17  &  25  &  1.08  &  SD \\
G291.90-0.013  &  291.90  &  -0.01  &  6.80  &  4.30  &  36.7  &  0.16  &  10  &  0.25  &  SD \\
G290.25-1.002  &  290.24  &  -1.04  &  17.8  &  10.2  &  18.8  &  0.21  &  19  &  0.58  &  SD \\
G292.26-0.679  &  292.25  &  -0.64  &  11.2  &  7.86  &  69.8  &  0.12  &  14  &  2.50  &  SD \\
G290.66+0.332  &  290.62  &  0.31  &  13.1  &  12.1  &  119.  &  0.29  &  31  &  6.75  &  NDA \\
G291.48-1.621  &  291.48  &  -1.62  &  4.85  &  4.85  &  0.00  &  0.45  &  10  &  0.66  &  NDA \\
G293.05-0.956  &  293.05  &  -0.95  &  16.0  &  16.0  &  0.00  &  0.18  &  31  &  2.87  &  NDA \\
G293.47-0.968  &  293.53  &  -0.94  &  28.0  &  9.98  &  136.  &  0.17  &  40  &  0.53  &  SD \\
G293.61-0.946  &  293.58  &  -0.97  &  9.23  &  8.65  &  0.00  &  0.34  &  10  &  1.00  &  SD \\
G293.66-1.223  &  293.62  &  -1.27  &  20.7  &  17.6  &  156.  &  0.15  &  37  &  0.37  &  NDA \\
G294.22-0.814  &  294.22  &  -0.81  &  7.97  &  4.38  &  50.6  &  0.22  &  10  &  0.66  &  NDA \\
G295.16-0.720  &  295.16  &  -0.74  &  24.7  &  8.65  &  14.0  &  0.19  &  35  &  2.88  &  SD \\
G294.47-0.482  &  294.47  &  -0.48  &  14.2  &  14.2  &  0.00  &  0.24  &  22  &  0.69  &  SD \\
G294.36-1.515  &  294.36  &  -1.51  &  9.36  &  9.36  &  0.00  &  0.13  &  13  &  1.16  &  SD \\
G295.86-0.143  &  295.85  &  -0.14  &  8.48  &  7.66  &  168.  &  0.20  &  20  &  0.81  &  SD \\
G297.26-0.969  &  297.26  &  -0.96  &  15.5  &  15.5  &  0.00  &  0.19  &  18  &  0.05  &  NDA \\
G298.74-0.345  &  298.70  &  -0.38  &  45.9  &  33.2  &  12.8  &  0.15  &  177  &  2.00  &  SD \\
G298.70+0.095  &  298.68  &  0.11  &  7.24  &  5.71  &  177.  &  0.22  &  13  &  0.62  &  SD \\
G299.06+0.188  &  299.06  &  0.18  &  13.4  &  13.4  &  0.00  &  0.26  &  21  &  0.61  &  SD \\
G297.96-0.515  &  297.96  &  -0.51  &  13.7  &  13.7  &  0.00  &  0.23  &  10  &  0.11  &  SCC \\
G299.34-0.297  &  299.33  &  -0.30  &  19.6  &  9.86  &  98.9  &  0.17  &  28  &  0.75  &  SD \\
G299.80+0.021  &  299.78  &  0.04  &  17.7  &  7.64  &  4.05  &  0.11  &  14  &  0.55  &  SD \\
G300.16-0.110  &  300.14  &  -0.10  &  10.3  &  6.66  &  0.00  &  0.34  &  16  &  1.66  &  SD \\
G300.09-0.384  &  300.08  &  -0.38  &  6.26  &  5.74  &  89.9  &  0.35  &  10  &  0.25  &  NDA \\
G300.20-0.353  &  300.21  &  -0.38  &  11.0  &  4.45  &  67.9  &  0.23  &  14  &  1.00  &  SD \\
G300.50-0.180  &  300.50  &  -0.18  &  13.1  &  10.4  &  68.3  &  0.29  &  29  &  1.41  &  SD \\
G300.65+0.067  &  300.66  &  0.11  &  21.7  &  12.7  &  111.  &  0.15  &  36  &  0.24  &  SD \\
G301.09+0.053  &  301.09  &  0.07  &  10.7  &  6.35  &  34.8  &  0.21  &  17  &  0.30  &  SD \\
G301.62-0.117  &  301.62  &  -0.11  &  12.1  &  5.50  &  143.  &  0.21  &  15  &  0.36  &  SD \\
G301.53+0.285  &  301.55  &  0.27  &  9.68  &  6.73  &  169.  &  0.13  &  10  &  0.42  &  SD \\
G301.78+0.269  &  301.81  &  0.25  &  13.3  &  6.47  &  60.0  &  0.20  &  17  &  0.21  &  NDA \\
G301.98+0.272  &  302.01  &  0.25  &  11.7  &  6.26  &  34.3  &  0.20  &  14  &  0.55  &  NDA \\
G301.97+0.717  &  301.99  &  0.70  &  34.9  &  13.1  &  146.  &  0.17  &  51  &  0.34  &  NDA \\
G302.06-0.009  &  302.04  &  -0.01  &  17.5  &  7.57  &  136.  &  0.27  &  31  &  1.58  &  SD \\
G303.04+0.174  &  303.04  &  0.18  &  6.70  &  6.00  &  89.9  &  0.31  &  10  &  0.66  &  SD \\
G302.60+0.401  &  302.60  &  0.41  &  12.6  &  10.5  &  175.  &  0.19  &  22  &  0.00  &  SD \\
G303.55-0.603  &  303.55  &  -0.60  &  21.7  &  21.7  &  0.00  &  0.20  &  48  &  1.66  &  SD \\
G304.13-0.390  &  304.13  &  -0.39  &  13.8  &  13.8  &  0.00  &  0.21  &  23  &  0.15  &  SD \\
G304.34-0.304  &  304.34  &  -0.30  &  9.50  &  9.50  &  0.00  &  0.22  &  15  &  1.50  &  SD \\
G304.05-0.538  &  304.05  &  -0.53  &  9.68  &  9.68  &  0.00  &  0.27  &  14  &  0.27  &  SD \\
G303.97-0.044  &  304.04  &  -0.05  &  15.9  &  7.34  &  0.00  &  0.25  &  17  &  0.70  &  SD \\
G304.72+0.307  &  304.72  &  0.26  &  32.2  &  21.3  &  134.  &  0.19  &  85  &  0.46  &  SD \\
G304.78+0.599  &  304.78  &  0.59  &  8.88  &  5.86  &  54.6  &  0.28  &  15  &  0.36  &  SD \\
G304.43+0.403  &  304.43  &  0.40  &  23.6  &  4.43  &  162.  &  0.20  &  25  &  0.13  &  SD \\
G304.92+0.596  &  304.92  &  0.59  &  9.67  &  5.23  &  100.  &  0.28  &  16  &  1.66  &  SD \\
G305.51+0.097  &  305.63  &  0.08  &  111.  &  47.0  &  24.5  &  0.14  &  466  &  1.08  &  SD \\
\hline
\multicolumn{10}{c}{
\tablefoot{
\tablefoottext{a}{Columns [1-6] are the same as in Sect. 4.1.}\\
\tablefoottext{b}{Column [7] is the surface density of clumps within the fitted ellipse in units of $\text{arcmin}^{-2}$.}\\
\tablefoottext{c}{Columns [8-9] correspond to the descriptions of Cols. [14-15] in Sect. 4.1.}\\
\tablefoottext{d}{Column [10] is a flag that determines its classification as a potential cluster candidate. No distance available (NDA) classifies sources without HDEs. Spread distance (SD) is a cluster to which no particular distance can be assigned as there are only a few clumps with HDEs and these HDEs show some spread. The final classification is SCC, which are cluster candidates that have a confused line of sight.}\\
}
}
\end{longtable}
\end{landscape}
\end{longtab}

   %*****************************************************************************
   %********************                                     ********************
   %********************              Results                ********************
   %********************                                     ********************
   %*****************************************************************************

\section{Results}

\subsection{Spatial distribution and cluster characterisation}

\subsubsection{Cluster candidate spatial distribution}

\begin{figure}
\resizebox{\hsize}{!}{\includegraphics{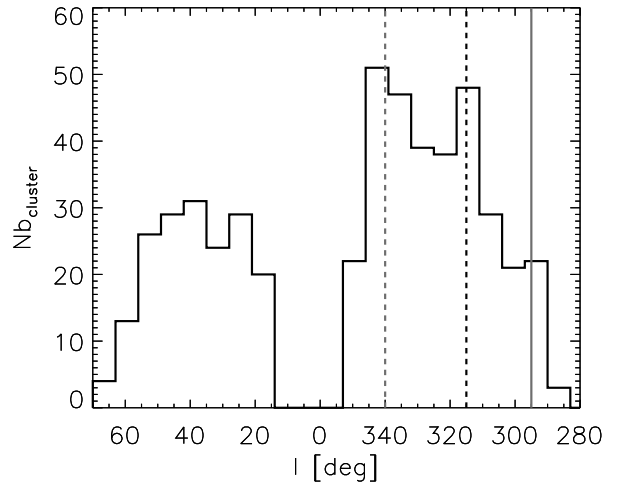}}
\caption{Longitude distribution of the 496 cluster candidates. The visible tangents of the spiral arms are located at $\emph{l}$=315deg (black dashed line). The tangent at $\emph{l}$=295 deg is marked by a grey line.}
\label{lonprofil}
\end{figure}

\begin{figure*}
\centering
\includegraphics[width=\textwidth,clip]{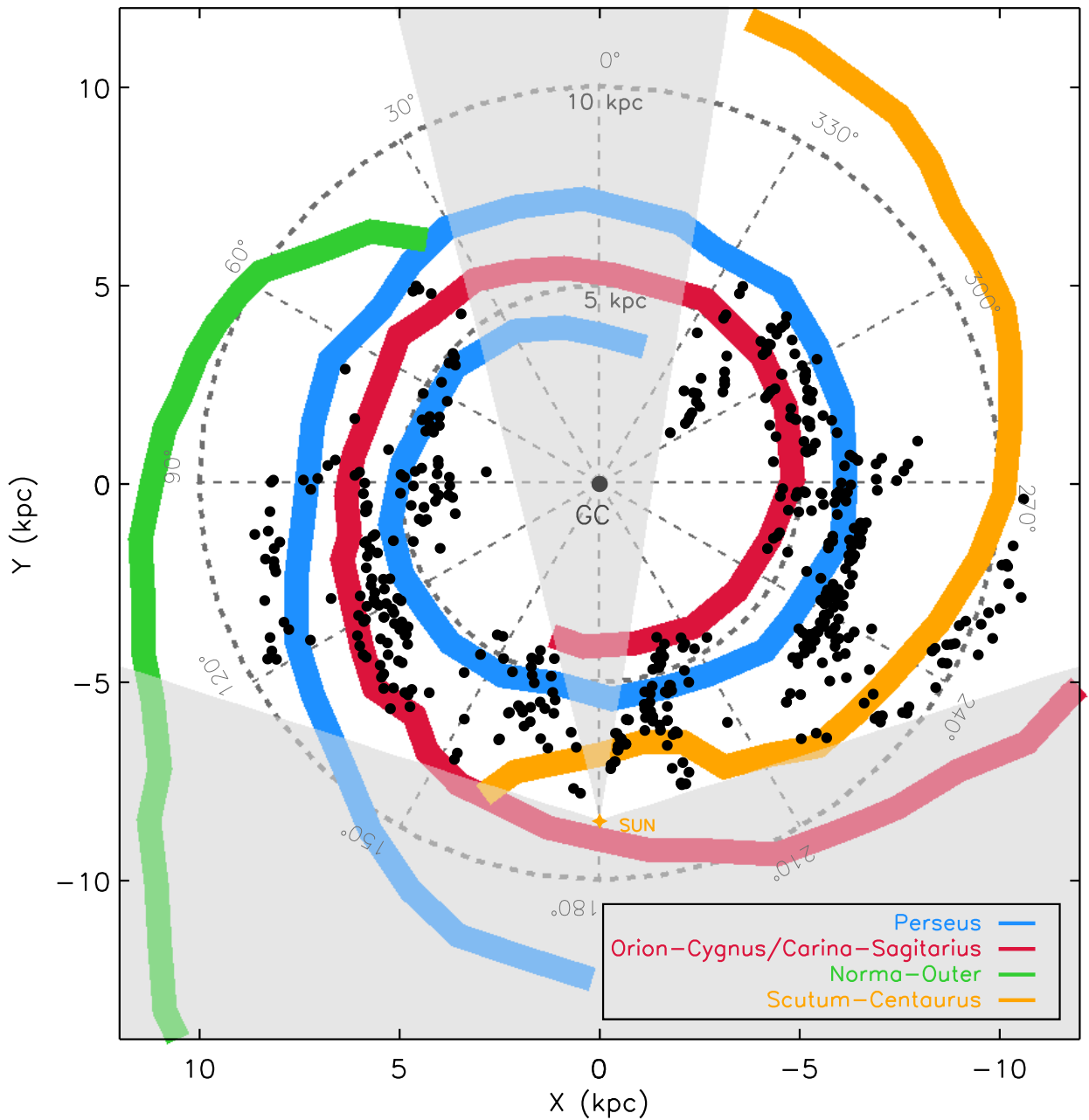}
\caption{Top-down view of the Galaxy using the HDEs of the cluster candidates. The spiral arm model is supplied by \citet{Englmaier2011}, with the different colour lines marking their locations. The plus (+) symbol represents the position of the Sun, with the grey dashed lines corresponding to Galactocentric distances. The black dots mark the location of the cluster candidates. The two grey shaded regions represent the regions where no cluster candidates could be found.}
\label{spiralarms}
\end{figure*}

The sources in this study are distributed between the first and fourth Galactic quadrants, with $\sim$35\%, 176 of the 496, found in the first quadrant. We split the two quadrants into two segments after visually inspecting the peaks in the longitude distribution of the cluster candidates, as displayed in Fig.~\ref{lonprofil}. The segments are [15,40] and [40, 60], and [320, 345] and [300, 320] for the first and fourth quadrants, respectively. We found 97, 72, 151, and 101 cluster quadrants in each of these segments, respectively. By using the HDEs, a 3D map can be produced, as displayed in Fig.~\ref{spiralarms}. The cluster candidates trace the spiral arms well, with the spiral arms from the study of \citet{Englmaier2011} using the Galactic distribution of molecular gas. The good agreement with the molecular-gas traced spiral arms is achieved because the clumps are very young and have not had time to migrate from the molecular clouds from which they formed. Unsurprisingly, a majority of sources are found associated with the tangents of spiral arms, with peaks in Fig.~\ref{lonprofil} in the fourth quadrant corresponding to the tangent at $\emph{l}$ = 315. Other tangents, at $\emph{l}$ = 30 and $\emph{l}$ = 295, are washed out of the sample, either by the number of cluster candidates at distances closer to the Sun or by the edge of the studied region, respectively.

\begin{figure}
\resizebox{\hsize}{!}{\includegraphics{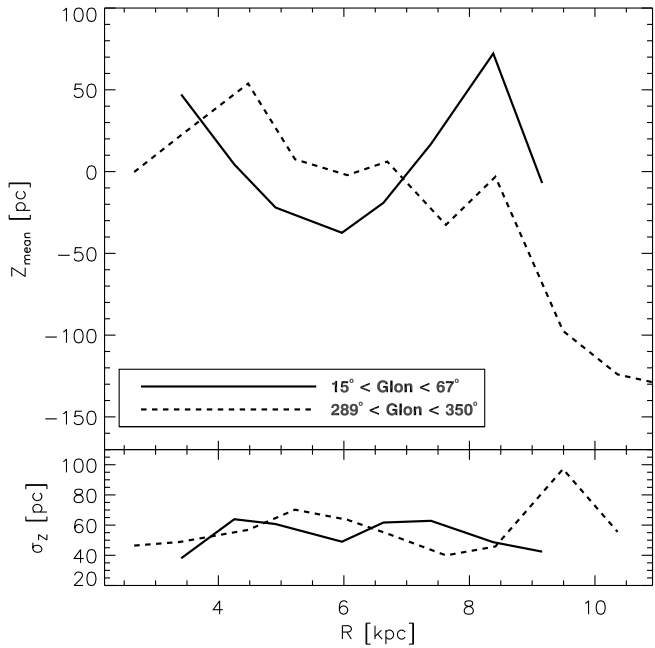}}
\caption{\textit{Top}: Distribution of the mean scale height, $Z_\text{mean}$, of cluster candidates as a function of Galactocentric radius. The black line corresponds to the cluster candidates in the first quadrant with the grey dashed line corresponding to the fourth quadrant. \textit{Bottom}: Distribution of the standard deviation, $\sigma_\text{Z}$, as a function of Galactocentric radius, $R$.}
\label{meanzdistrib}
\end{figure}

We also analysed the scale height, Z, of the cluster candidates from the Galactic Plane using the expression

\begin{equation}
Z = D_\text{peak} \sin(b)
\label{equaZ}
,\end{equation}
where $D_\text{peak}$ is the heliocentric distance of the cluster candidates. The HDE we used is the peak value of the cluster distance histogram. The Galactic latitude represents the central position of the cluster candidate. We also calculated the Galactocentric radius of each source using the relation

\begin{equation}
R = \sqrt{R_\text{0}^2 + D_\text{peak}^2 - 2R_\text{0}D_\text{peak}\cos(l)}
\label{equaR}
,\end{equation}
where $R_\text{0}$ represents the Galactocentric distance of the Sun, set at 8.5\,kpc. By calculating the mean scale height in Galactocentric radius bins of 0.1\,kpc, we computed the profile of $Z_\text{mean}$ over the first and fourth quadrants; this
is displayed in Fig.~\ref{meanzdistrib}. We show in the lower panel of Fig.~\ref{meanzdistrib} the standard deviation, $\sigma_\text{Z}$, in each bin along R. We compared the $Z_\text{mean}$ profile to that of \citet{Paladini2004}, who observed the distribution of \ion{H}{ii} regions in the entire Galactic Plane. The profile in this study matches that of \citet{Paladini2004} in the fourth quadrant, with a break at $\sim$ 7\,kpc. The first quadrant shows a decrease in $Z_\text{mean}$ until $\sim$ 6\,kpc, at which an increase is observed (also observed by \citet{Paladini2004}) until another decrease at 8.5\,kpc. However, the range of $R$ in the first quadrant is much shorter than that of the fourth, with no access to the sources at R greater than 10\,kpc, where \citet{Paladini2004} observed an increase. The difference between the two quadrants at Galactocentric radii greater than 7\,kpc is most likely due to the warp of the Galactic Plane \citep[see][]{Burton1988}. The warp is observed in \ion{H}{I} \citep{Henderson1982} as well as molecular clouds, OB stars, and the stars traced by 2MASS. An exhaustive list of the main components of the warp can be found in \citet{Reyle2008}. The warp mostly occurs at Galactocentric radii larger than the solar circle. The cluster candidates lie in a thin, asymmetric disk in a range of Z from -248 to 134\,pc. The asymmetry is due to the location of the Sun above the Galactic Plane \citep[e.g.][]{Brand1993,Reed1997,Reed2006}, as this would have more effect on the nearby sources. The mean value of Z in the entire study is -6.2\,pc, again because of the position of the Sun.

Our data suggest the presence of the Galactic warp, but its amplitude is quite ambiguous. Similar results are also found in the star distribution, where the warp is hardly recognizable \citep{Marshall2006,Reyle2008} with respect to the gas distribution. The study by \citet{Paladini2004}
 shows a flare at the scale height after 10\,kpc, which cannot be observed here as shown with $\sigma_\text{Z}$ in the lower panel of Fig.~\ref{meanzdistrib}. However, our data do not exclude the flaring in the outer disk that is generally traced by the star distribution \citep{Derriere2001}.

\begin{figure}
\centering
\begin{subfigure}[b]{9cm}
   \resizebox{\hsize}{!}{\includegraphics{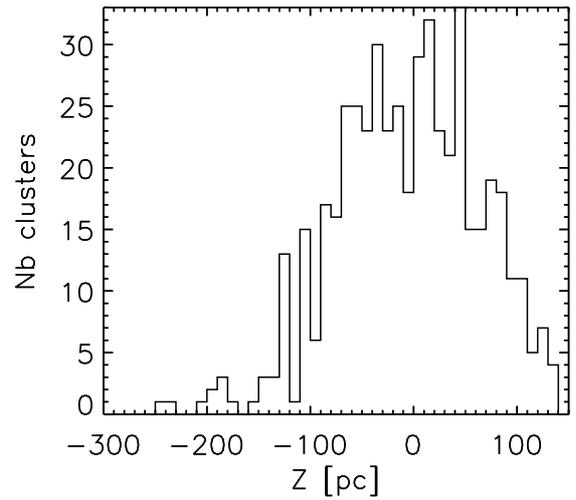}}
   \label{zdistrib}
\end{subfigure}
\quad
\centering
\begin{subfigure}[b]{9cm}
   \resizebox{\hsize}{!}{\includegraphics{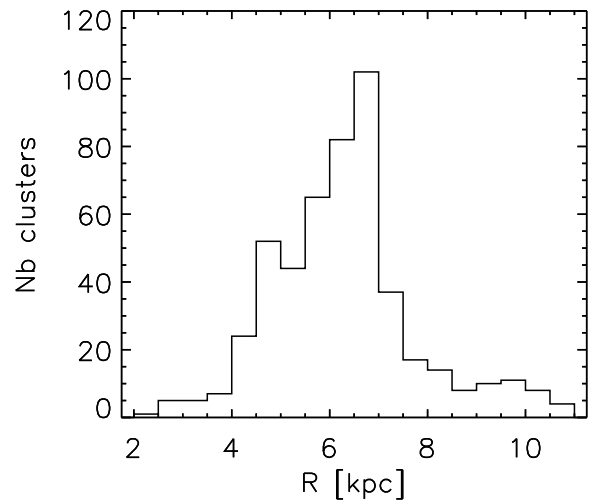}}
   \label{rdistrib}
\end{subfigure}
\caption{\textit{Top}: Histogram of scale heights of the cluster candidates. The mean value is -6.2\,pc. \textit{Bottom}: Distribution of cluster candidate Galactocentric radii. The peak is at $\sim$ 6.5\,kpc, a value also observed by \citet{Paladini2004}.}
\label{distrib}
\end{figure}

\subsubsection{Spatial distribution of clumps in clusters}

We used the parameter $Q$ introduced by \citet{Cartwright2004} to distinguish between centrally concentrated cluster candidates and those with a fractal distribution. The parameter is defined as

\begin{equation}
Q = \frac{\bar{m}}{\bar{s}}
,\end{equation}
where $\bar{m}$ is the normalised mean branch length of the sub-tree of the cluster candidates, calculated from $(N_\text{clumps}A)^{0.5}/(N_\text{clumps}-1),$ where $N_\text{clumps}$ is the total number of clumps inside each cluster candidate and A is the area of the cluster. $\bar{s}$ is the mean distance between clumps divided by the cluster candidate radius. Using simulated clusters, \citet{Cartwright2004} identified fractal clusters as having a $Q$ parameter value $\le0.8$, whereas $Q>0.8$ corresponds to concentrated clusters. Clusters with $Q=0.8$ correspond to clumps with a random distribution, that is, unclustered clumps. Subsequent work of \citet{Cartwright2009b} found that clusters in the range $0.75\le Q \le0.85$ could also follow a concentrated or fractal distribution. We found a range from $Q=0.4$ to $Q=0.84$ with 98.6\% of cluster candidates below $Q=0.8$ and 90\% below $Q=0.75$. This means that the cluster candidates follow a fractal distribution, as defined by the Q parameter of \citet{Cartwright2004}, as does the ISM \citep{Combes2000}. This range is similar to one found by \citet{Parker2015c} in hydrodynamical simulations.

\citet{Cartwright2009a} found that the elongation of the cluster, the ratio between the semi-major and minor axes, could affect the value of Q for elongations above 3. Here only 40 cluster candidates have an elongation above this value, and no correlation between the two were found, therefore we decided against a correction to Q. The elongation could represent a filamentary distribution, therefore some cluster candidates with $Q=0.8$ do not necessarily mean a random spatial distribution. Furthermore, \citet{Cartwright2004} studied star clusters that may have a different distribution and may not present filamentary structures compared to the young sources investigated here.
Other effects, typically observational limitations and statistical biases, can affect the Q parameter as developed in \citet{Bastian2009b}. The cluster candidates suffer from a lack of statistics because of the low number of clumps (88\% of cluster candidates have $\leq 30$ clumps) compared to the star clusters of \citet{Cartwright2004} ($\geq 100$). This could affect the profile estimation of the cluster candidates. For example, a cluster with a random distribution but with a low statistic can have a fractal profile. By restraining the number of cluster candidates to those with the greater number of clumps ($\geq 30$ clumps), we observe that the values of the Q parameter remain in the same range, with a range from $Q=0.40$ to $Q=0.77$.\\

Two more metrics can be used to determine whether the cluster candidates are significant, or in other words, that their shapes are not random. The first of these is that of \citet{Campana2008},

\begin{equation}
g_\text{k} = \frac{\bar{\Lambda}}{\bar{\Lambda}_\text{k}}
,\end{equation}
where $\bar{\Lambda}$ is the mean branch length and $\bar{\Lambda}_\text{k}$ is the mean branch length of the k-th cluster. We find a range of $g_\text{k}$ from 0.95 to 3.02 with a peak at 1.5. The second metric is that of \citet{Massaro2009}, which uses the relation between the number of clumps inside each cluster and $g_\text{k}$,

\begin{equation}
M_\text{k} = N_\text{clumps} g_\text{k}
,\end{equation}
where $N_\text{clumps}$ is the total number of clumps in the k-th cluster. This metric, called \textit{\textup{magnitude}}, allows us to compare clusters with all ranges of numbers of clumps. $M_\text{k}$ has values from 9.5 to 380 and 54\% of cluster candidates have $M_\text{k} > 20$. The peak of this distribution for our cluster candidates, which is 15, corresponds to a value of $g_\text{k} = 1.5$ with $N_\text{clumps} = 10$. The two metrics give results indistinguishable from each other and are consistent with the studies that reported the definitions in the first place.\\

\subsection{Clump mass functions}

We compared the masses of clumps found in the 496 cluster candidates to those that are considered isolated. Any clump that is connected only by branch lengths greater than the cut-off branch length was considered to be isolated. This gives a total of 4\,752 isolated clumps compared to 8\,248 clustered clumps. The masses are computed with

\begin{equation}
M = \frac{F_\nu D^2}{\Omega k_\nu B(T,\nu)}
\label{equaM}
,\end{equation}
with the masses normalized with the quantity $D_\text{peak}^{2}/D^{2}$ by taking the ratio of the peak distance in the cluster distance histogram compared to the individual source distance.

These masses allow for the clump mass function (CMF) for the clustered and isolated clumps to be compared. The relation outlined in Eq.~\ref{equaIMF} was assumed \citep{Kroupa1993}. $\alpha$ is linked to the Salpeter slope \citep{Salpeter1955} by the relation $\alpha = \Gamma + 1$,

\begin{equation}
\Phi = \frac{dn}{dm} \propto m^{-\alpha} \, \text{with} \, \alpha = \left\{
    \begin{array}{lll}
       \alpha_1, \, M_0 < M < M_1 \\
       \alpha_2, \, M_1 < M < M_2 \\
       \alpha_3, \, M_2 < M
    \end{array}
\right.
\label{equaIMF}
.\end{equation}
The bin size used in this study does not significantly alter the values of $\alpha$ as the number of clumps used is $N > 500$ \citep{Rosolowsky2005}. A logarithmic bin of mass 0.1 was used for all CMFs, corresponding to the criterion of $\text{W} = 2(IQR)/\sqrt[3]{N,}$ where W is the bin width, IQR is the interquartile range, and N is the number of sources in the whole CMF \citep{Freedman1981}.

\begin{figure*}
\resizebox{\hsize}{!}{\includegraphics{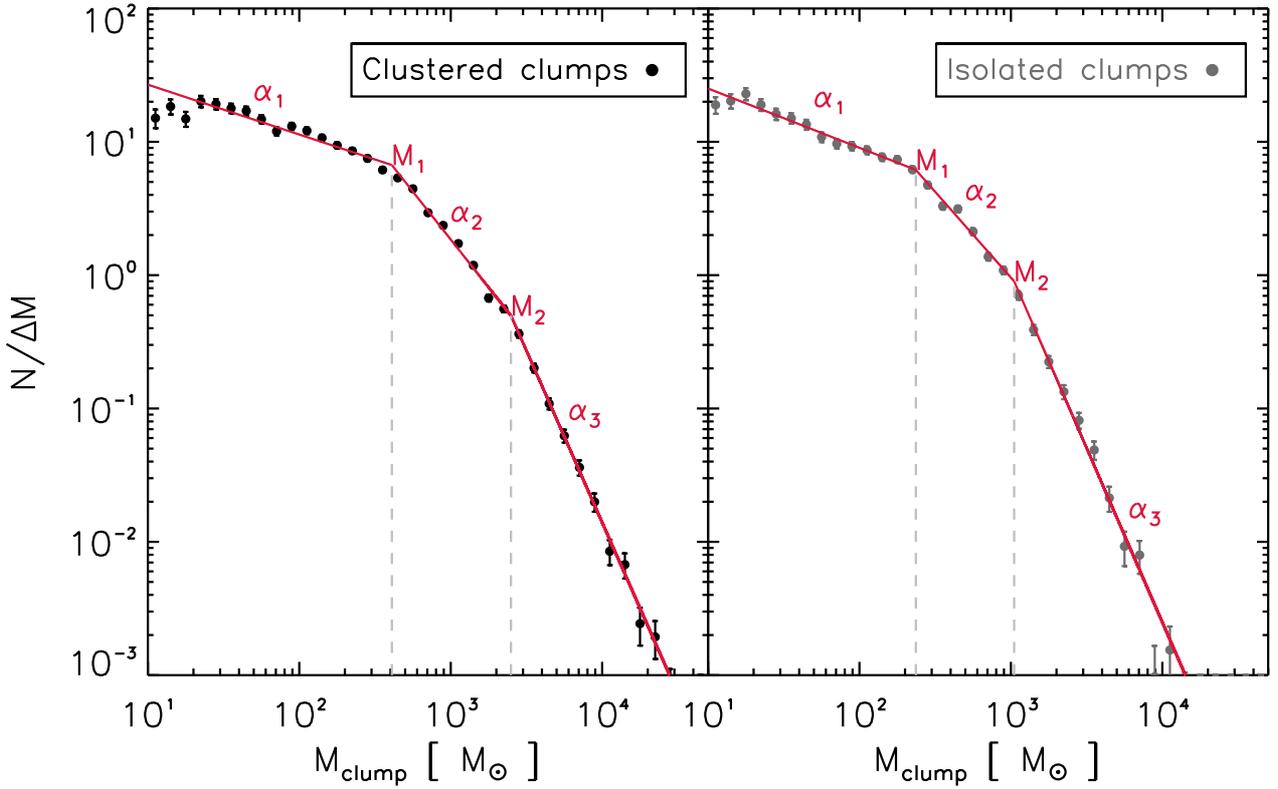}}
\caption{CMFs of clustered (left panel) and isolated (right panel) clumps. The masses of the clumps are computed using Eq.~\ref{equaM} and the ratio $D_\text{peak}^{2}/D^{2}$, where $D_\text{peak}$ is the distance of the cluster candidates and $D$ the individual distance of clumps. Only clumps with HDEs are used. The two plots correspond to the function described in Eq.~\ref{equaIMF}. The red lines correspond to the fitted segments of the CMF, which provide the lopes $\alpha_1$, $\alpha_2$ and $\alpha_3$ as well as the break masses $M_1$ and $M_2$.}
\label{imfclustiso}
\end{figure*}

Figure~\ref{imfclustiso} shows the CMFs for clustered (left panel) and isolated (right panel) clumps. The error bars correspond to the 1$\sigma$ Poisson errors, $\sigma = \sqrt{N}/\Delta M$. The errors on the individual clump masses are not considered because the main error source is the distance uncertainty. A three-segment power law was fit to each CMF, with the form of Eq.~\ref{equaIMF}. These three segments were compared with those of Kroupa's IMF \citep{Kroupa2001,Kroupa2002}. The lower bound is assumed constant in the analysis and equal to $10 \, M_{\sun}$. The fits provide values for the three quantities $\alpha_1$, $\alpha_2$ , and $\alpha_3$, as well as the mass breaks, $M_1$ and $M_2$. Table~\ref{cmfvalue} displays the fitting results for clustered and isolated clumps. These slopes are consistent both with each other and with the Kroupa IMF. The sample was also split into the two quadrants, and the clustered and isolated CMFs were reproduced. The results are also consistent with the Kroupa IMF. Table~\ref{cmfvalue} also contains the fitting results for these CMFs. The figures of the first and fourth quadrant CMFs can be found in Appendix A (Figs.~\ref{isoclustfirstqd} and~\ref{isoclustfourthqd}).

In all cases the break masses $M_1$ and $M_2$ are much greater in clustered than in isolated clumps.

\begin{table*}
\caption{CMF fitting results for clustered and isolated clumps in the whole sample, first and fourth quadrants.}
\label{cmfvalue} 
\centering
\begin{tabular}{c@{\hskip 0.3in} | c@{\hskip 0.3in} c@{\hskip 0.3in} c@{\hskip 0.3in} c@{\hskip 0.3in} c@{\hskip 0.3in}}
\hline\hline 
Clump types & $\alpha_1$ & $\alpha_2$ & $\alpha_3$ &  $M_1 \, (M_{\sun})$ & $M_2 \, (M_{\sun})$ \T\B \\
\hline
Clustered clumps & 0.37 $\pm$ 0.02 & 1.43 $\pm$ 0.03 & 2.57 $\pm$ 0.07 & 409 & 2,500 \T \\
Isolated clumps & 0.44 $\pm$ 0.02 & 1.29 $\pm$ 0.05 & 2.61 $\pm$ 0.04 & 236 & 1,050 \B \\
\hline
\multicolumn{6}{c}{}\\
\multicolumn{6}{c@{\hskip 0.3in}}{First quadrant}\\
\hline
Clustered clumps & 0.15 $\pm$ 0.05 & 1.29 $\pm$ 0.06 & 2.60 $\pm$ 0.14 & 324 & 2,080 \T \\
Isolated clumps & 0.31 $\pm$ 0.06 & 1.35 $\pm$ 0.07 & 2.63 $\pm$ 0.09 & 216 & 1,010 \B \\
\hline
\multicolumn{6}{c}{}\\
\multicolumn{6}{c@{\hskip 0.3in}}{Fourth quadrant}\\
\hline
Clustered clumps & 0.46 $\pm$ 0.02 & 1.55 $\pm$ 0.04 & 2.56 $\pm$ 0.11 & 483 & 2,880 \T \\
Isolated clumps & 0.53 $\pm$ 0.03 & 1.24 $\pm$ 0.08 & 2.35 $\pm$ 0.10 & 251 & 960 \B \\
\hline
\end{tabular}
\end{table*}

\begin{figure}
\resizebox{\hsize}{!}{\includegraphics{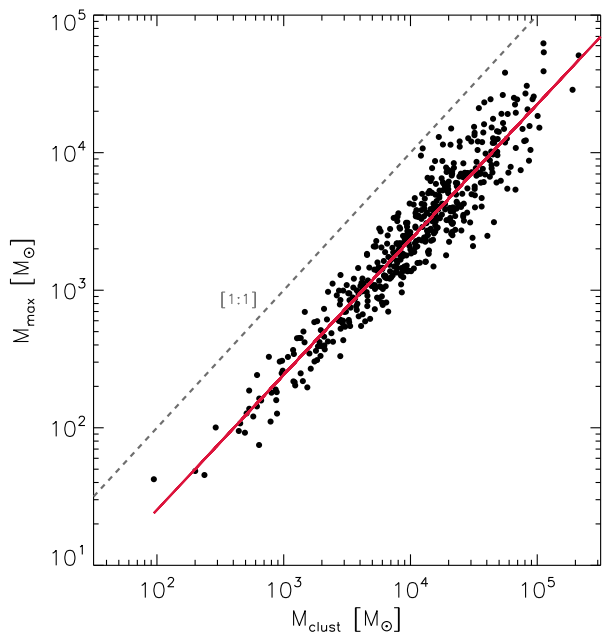}}
\caption{Highest mass clump in each cluster candidate as a function of total cluster mass. The grey dashed line is the 1:1 line and a highest value for the highest mass clump. The red line is the linear fit to the distribution.}
\label{massclusterbigmass}
\end{figure}

A correlation is found between the highest mass clump in each cluster candidate and the total cluster mass, as shown in Fig~\ref{massclusterbigmass}. The dashed line corresponds to the upper limit of the maximum mass, that is, the cluster mass. The red line is the linear fit to the distribution, $M_\text{cloud,max} = \zeta M_\text{cluster}^\eta$ with $\zeta = 0.27 \pm 0.04$ and $\eta = 0.98 \pm 0.02$. This correlation remains the same when the sample is split into heliocentric distance bins. \citet{Kirk2012} and \citet{Weidner2010} observed the same correlation for YSO groups, but with a flatter slope of $\eta \approx 0.5$. This suggests that different mechanisms act at these scales in producing and fragmenting the clumps.

   %*****************************************************************************
   %********************                                     ********************
   %********************             Discussion              ********************
   %********************                                     ********************
   %*****************************************************************************

\section{Discussion}

\subsection{Completeness of the cluster candidate catalogue}

The MST is a powerful method for finding over-densities in images, especially when over-densities are composed of filamentary structures. The branch-length cut-off, $\Lambda_\text{cut}$, can be determined by different methods. Here, the method described in \citet{Koenig2008} and \citet{Gutermuth2009} was used. Other methods include that of \citet{Battinelli1991}, who defined $\Lambda_\text{cut}$ as the branch length that corresponds to the maximum number of over-densities in a sample. \citet{Maschberger2010} chose the branch length cut-off such that the properties of over-densities found correspond to the ones selected by eye.

The drastic drop in the number of over-densities (1,633) found compared to cluster candidates (496) gives an idea of the line-of-sight projection effect that is stronger in the inner Galaxy. When the accuracy of the heliocentric distance estimates also improves, the number of cluster candidates  increases and may allow for MST to be computed in 3D. The heliocentric distances are also a great source of uncertainty in the computation of clump mass
that may affect the slopes of the CMF. A greater number of heliocentric distances will also increase the source numbers and reduce the Poisson errors in the CMF slope computations, which are the largest source of uncertainty in that analysis.\\

\begin{figure}
\resizebox{\hsize}{!}{\includegraphics{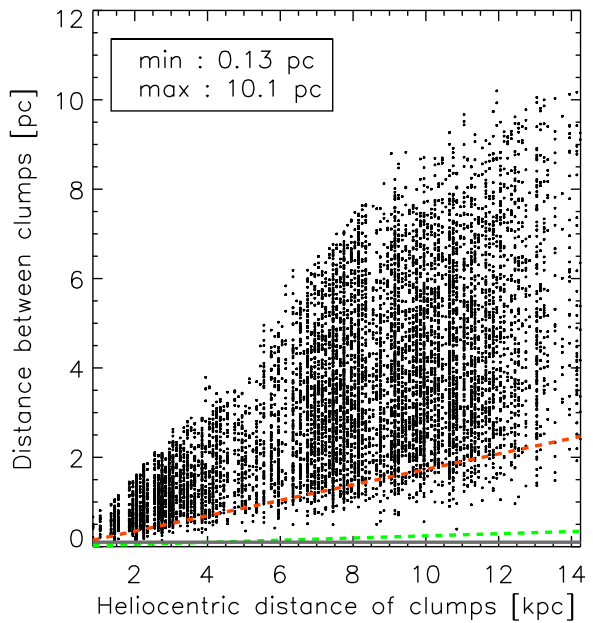}}
\caption{Distribution of the distance between clustered clumps in cluster candidates. The green dashed line represents the linear diameter of the beam at 70~$\mu m$. The orange dashed line corresponds to the same size, but at 500~$\mu m$. The grey line marks the Jeans length (0.1~pc).}
\label{distribdistclumps}
\end{figure}

The fractal distinction of the cluster candidates implies that the clumps are gravitationally bound to their host cluster, but also that a non-random sub-clustering distribution can be observed. This means that sources are probably more spaced than in the case of a centrally concentrated cluster. We can investigate the physical distance between clumps and a possible bias by comparing our results with the typical Jeans length of clumps (Eq.~\ref{equaJeans}), which characterises the ability of a sphere to collapse, as well as with the resolution at the extreme wavelengths 70~$\mu m$ and 500~$\mu m$,

\begin{equation}
\lambda_\text{j} = \sqrt{\frac{15 k_\text{B} T}{4 \pi G \mu n m_\text{p}}}
\label{equaJeans}
,\end{equation}
where $k_\text{B}$ is the Boltzmann constant, T is the temperature of the cloud, n is the mean volume density of the cloud, $m_\text{p}$ is the proton mass, and $\mu$ is the mean molecular weight. We considered temperatures of 10, 20, and 30\,K with a mean volume density of $10^5$ particles per $\text{cm}^{3}$ and an average mass per particle $\mu m_p = 4\times10^{-27}$~kg, assuming a cloud composed of 80\% hydrogen and 20\% helium and $\mu \approx 2.5$. These values give a Jeans length of 0.05\,pc to 0.1\,pc. This length would be larger if we considered a cloud with magnetic fields and under external pressure. Figure~\ref{distribdistclumps} shows the relation of projected distances (in pc) computed from the MST branches $\Lambda$ characterising the angular distances between clustered clumps, with the HDE of the clumps taken equal to the mean distance of the cluster candidates. The lowest value, $\approx 0.13$~pc, is comparable to the highest value of the Jeans length. Distances computed in three dimensions are greater than the projected distances and increase the difference with the Jeans length. We also plot the linear diameter of the beam at 70~$\mu m$ (in green) and 500~$\mu m$ (in orange) in the heliocentric distance range that covers our sample as well as the highest value of the Jeans length (0.1~pc, in grey). All physical distances between clumps are greater than the 70~$\mu m$ beam size and more than 90\% are greater than the 500~$\mu m$ beam size. This means that the clumps are separated by much more than the imprints that each of them leaves on the fragmented cloud. The PACS and SPIRE resolutions introduce a bias by not allowing us to detect clumps with a spacing shorter than the Jeans length, as has been mentioned in \citet{Billot2011}.\\

\begin{figure*}
\centering
\begin{tabular}{cc}
\includegraphics[width=0.45\textwidth]{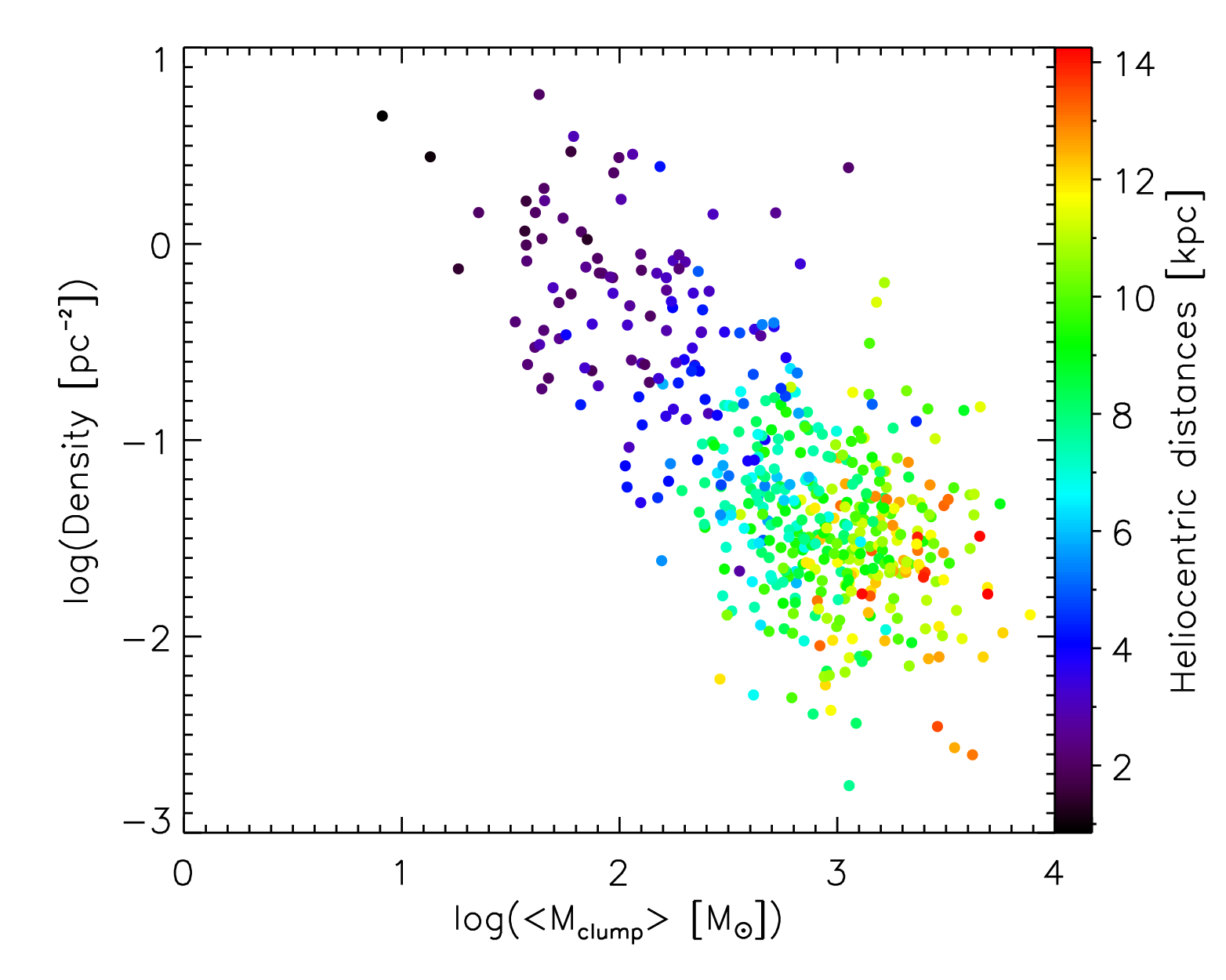} & \includegraphics[width=0.45\textwidth]{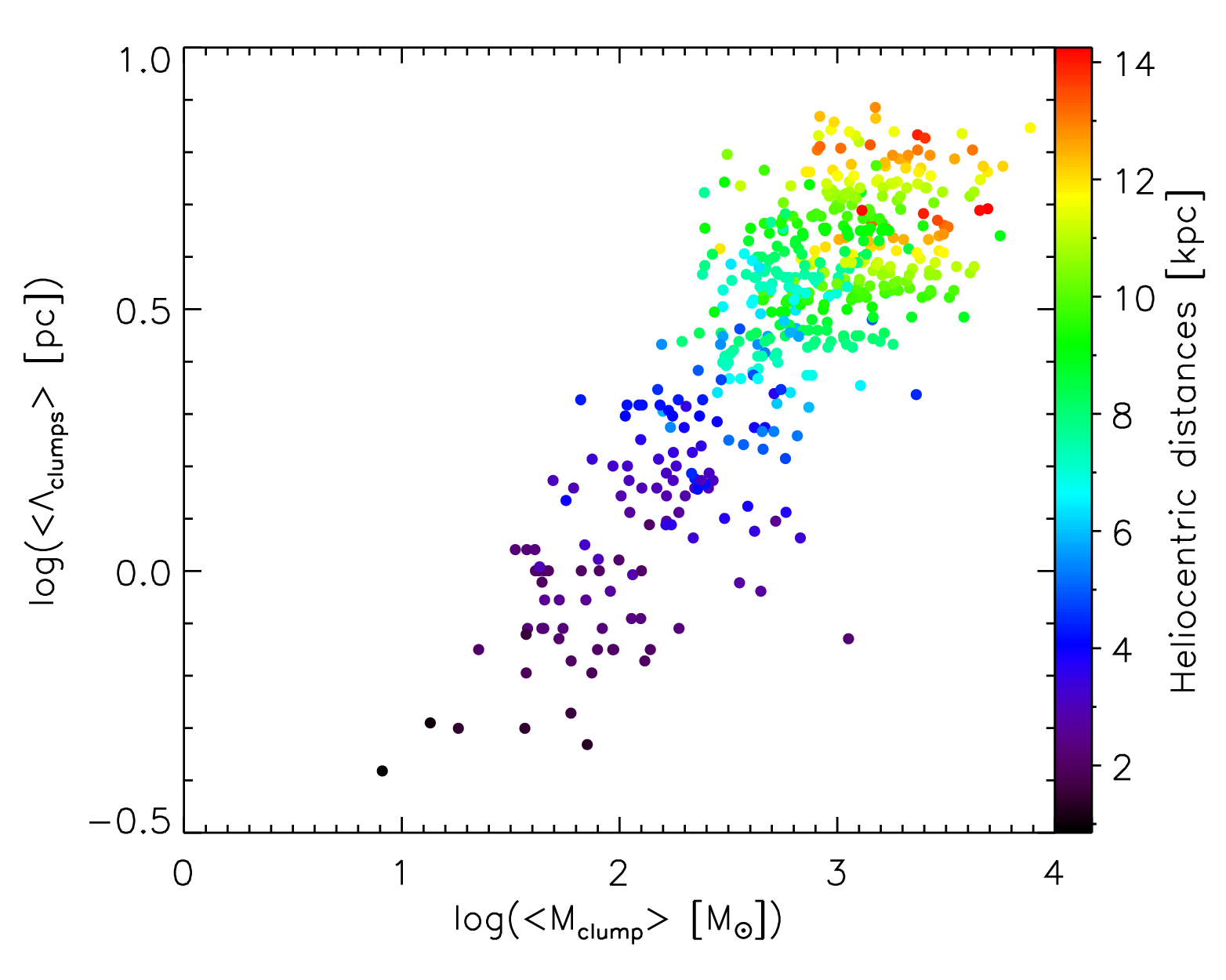}\\
\includegraphics[width=0.45\textwidth]{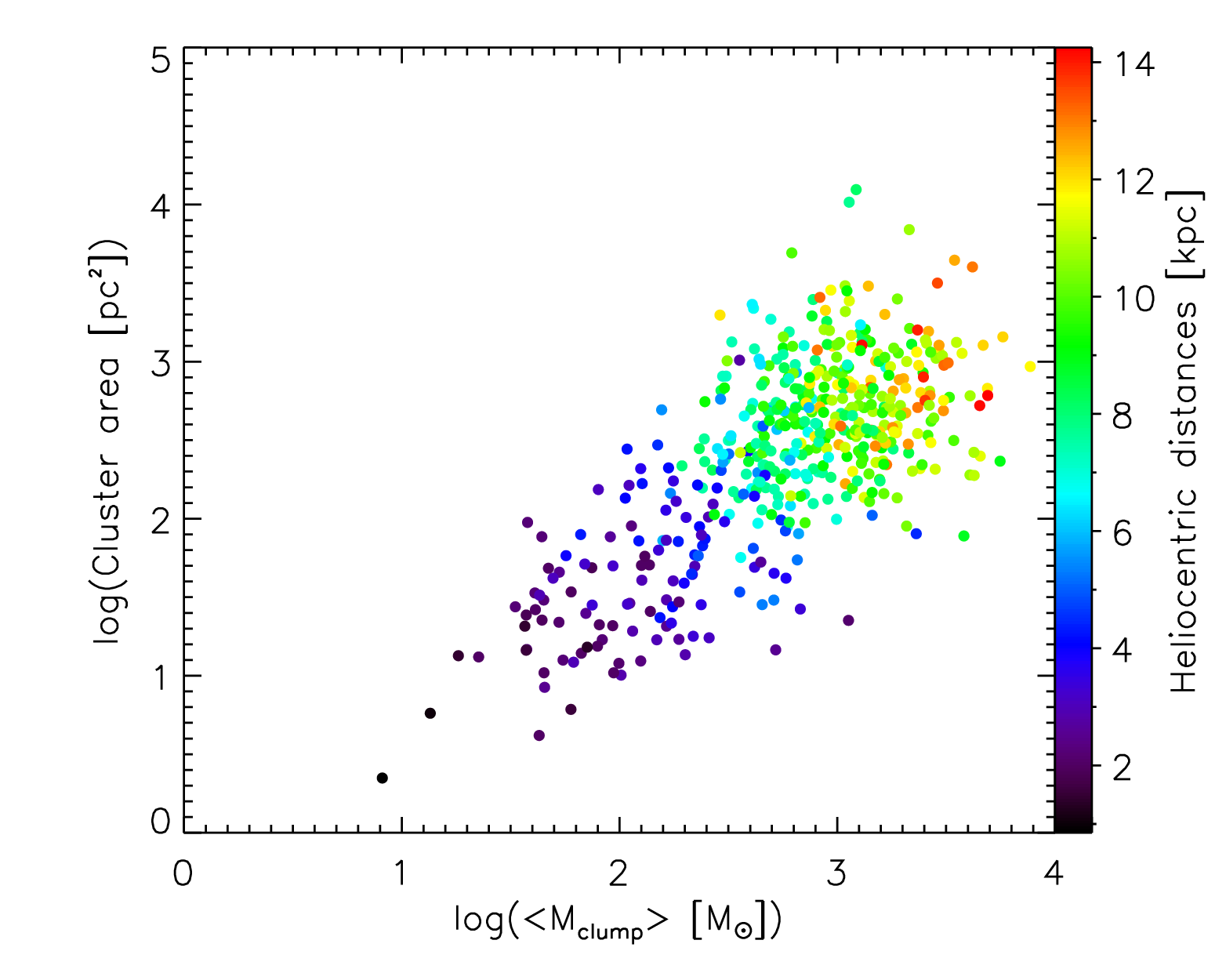} & \includegraphics[width=0.45\textwidth]{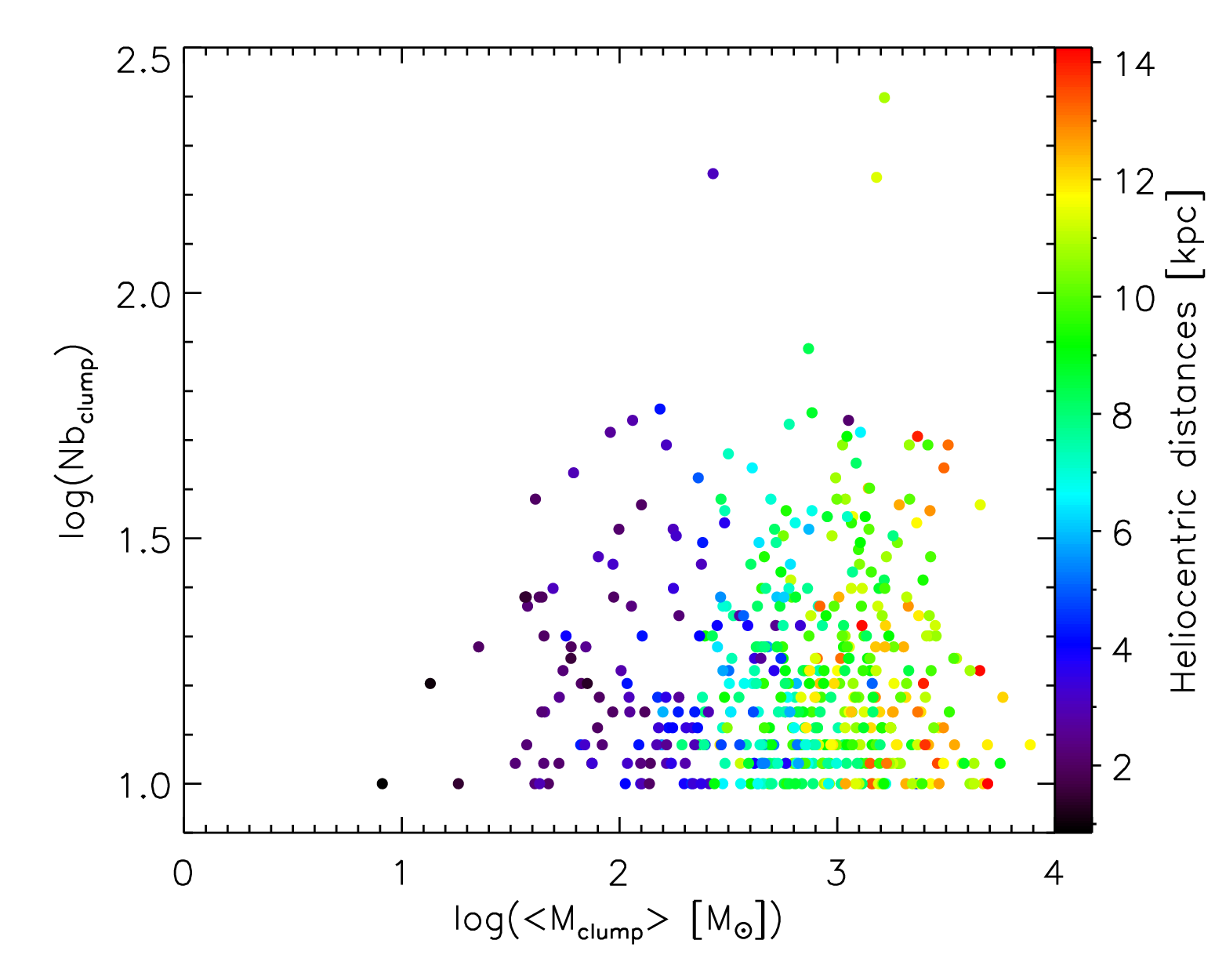}\\
\end{tabular}
\caption{Cluster candidate properties as a function of mean clump mass. Top left: cluster density versus mean clump mass. Top right: mean MST clump length versus mean clump mass. Bottom left: cluster area versus mean clump mass. Bottom right: number of clumps versus mean clump mass. The colour bar corresponds to the heliocentric distance of the cluster candidate.}
\label{cormasschar}
\end{figure*}

Figure~\ref{cormasschar} shows different properties of cluster candidates as a function of mean clump mass within the cluster candidate. The cluster density, mean distance between clustered clumps, cluster area, and number of clumps per cluster candidate are compared to the mean clump mass. The colour bar corresponds to the heliocentric distance of the cluster candidate. There is a strong correlation with mean clump mass and all the properties, except for the number of clumps. The cluster density, (Fig.~\ref{cormasschar}a), shows that the lower the cluster density, the higher the mean clump mass. Figures~\ref{cormasschar}a and~\ref{cormasschar}b show a positive correlation with mean clump mass. Figure~\ref{cormasschar}b shows a break at around 6\,kpc, which implies that beyond this distance, there is no significant increase in the distance between clumps. We are probably affected by a bias on the clump mass that also introduces another bias on the physical distance between clumps (mentioned above). As we are more sensitive to bright and massive clumps, it is possible that at a certain heliocentric distance we are not able to detect low-mass clumps. We can roughly estimate the critical distance for a typical mass by using the completeness limit of the single-band catalogues \citep{Molinari2016} and by inverting Eq.~\ref{equaM}, 

\begin{equation}
D = \sqrt{\frac{M k_\nu \Omega B(T, \nu)}{F_\nu}}
,\end{equation}
where $k_\nu = k_{ref} \left( \frac{\nu}{\nu_ref} \right)^2$ is the opacity law characterising the dust emissivity with $k_{ref} = 0.1~\text{cm}^2~\text{g}^{-1}$ at $\nu_{ref} = 250 \mu m$ \citep{Hildebrand1983}. The black-body law is computed at a temperature of 15~K. $\Omega$ is the solid angle of a source taken with an angular size of 20\arcsec here. The completeness limits are roughly estimated to 0.7, 1.5, 2.0, 2.0, and 3.0 Jy for 70, 160, 250, 350, and 500~$\mu m$, respectively. The result is that beyond 6-7~kpc, it becomes difficult to detect clumps with a mass lower than $1000~M_\odot$ in each band, which also corresponds to the break in Fig.~\ref{cormasschar}b.

\subsection{Discussion of the CMF and relation with IMF}

The Hi-GAL physical property catalogue provides a mass for clumps without HDEs, using a distance of 1\,kpc. This catalogue is a preliminary physical property catalogue since at the time of this analysis not all the distances were computed. In our analysis, we have converted the mass of clustered clumps without HDEs to the mass corresponding to the distance of their respective cluster candidate. To ensure that the HDEs do not alter the CMFs, we computed the CMFs with these distances unchanged and the whole sample. The properties, break masses, and slopes are not changed, and the results are presented in Appendix A (Fig.~\ref{masschangeclust}).

We did not split the clumps according to their evolutionary stages. The determination of the pre-stellar and proto-stellar stages, as explained in the introduction, produced a large number of pre-stellar clumps, and the possibility of misidentifying the criterion meant that we preferred to refrain from analysing this at present.

As outlined in the previous section, the farthest clumps, that is, those with distances greater than $\sim$ 5\,kpc, dominate the high-mass end of the CMF. However, as no bump that is due to the blending of lower mass sources with the highest mass clumps is observed in the high-mass end of the CMF \citep{Moore2007,Reid2010}, we can assume that clumps at large distances do not alter the CMF and that these clumps are part of the same distribution.

The results in Table~\ref{cmfvalue} suggest that the CMF follows Kroupa's IMF. The slopes of the Kroupa IMF are $\alpha_1=0.3$, $\alpha_2=1.3$ and $\alpha_3=2.3$ \citep{Kroupa2001}. The values are consistent within 3-$\sigma$. As explained in Sect. 5.2, we neglected the errors on the mass calculations, which are mainly due to the distance estimation and are conservatively estimated at 20\% \citep{Rosolowsky2005}.

The CMF and IMF seem to be consistent, which indicates that the IMF is set at the clump formation stage. Studying the ATLASGAL\footnote{\citet{Schuller2009}} clumps, \citet{Wienen2015} found a similar result in several embedded clusters composed of cores. \citet{Simpson2008} found that the core mass function (CoMF) also maps the IMF in the Ophiuchus cloud L1688. \citet{Bastian2010} suggested that the IMF was invariant across environment, so that if the IMF does follow the CoMF, then any variations in the CoMF may indicate changes in the stellar IMF. However, \citet{Michel2011} found that the CoMF does not necessarily follow the IMF in the densest groups of cores. \citet{Hatchell2008} also showed that the pre-stellar core mass function could be steeper than the IMF when the most massive pre-stellar cores fragment to form several less massive proto-stellar cores.\\
As most clumps are composed of several pre- and/or proto-stellar cores, this mix as well as the mix of pre- and proto-stellar clumps could easily change the CMF slope. For this reason, we have to be cautious that the link between the CMF and the IMF is probably less evident and the IMF-like profile of the CMF could suffer from observational effects (distance and mass estimation, resolution and mix of different star formation regions).\\
As the scanned mass range is wide, we have to be aware of a possible truncation in the mass distribution that is due to different types of object. Beyond 1000~$M_{\sun}$, clumps might be considered as molecular clouds (MC) and giant molecular clouds (GMC) for $M > 10^4~M_{\sun}$. Considering this difference, the slope of the molecular cloud mass function (MCMF), which corresponds to the highest parts of the CMF (as shown in Fig.~\ref{imfclustiso}), agrees with the Kroupa IMF slopes. The second slope $\alpha_2$ (see Table~\ref{cmfvalue}) also agrees with the slope found by \citet{Klessen2000} and \citet{Elmegreen1985}, $dN/dM \approx M^{-1.5}$ , who used different simulations to investigate the fragmentation of the molecular clouds. However, the slope $\alpha_3$ that characterised the highest part of the CMF is steeper than these values. \citet{Tsuboi2012} and \citet{Tsuboi2015} showed that the MCMF can vary close to the Galactic centre and corresponds to the $\alpha_3$ values in our study.\\

When we still consider all clumps as the same type of object and assuming the link between the CMF and the IMF, there exists a mass coefficient between clumps and stars. This coefficient is related to the star-formation efficiency (SFE) as well as to the multiplicity factor that characterises the number of stars formed in each clump, $n_*$. The relation between the turnover masses of the IMF and CMF is then $M_\text{IMF} = (\epsilon/n_*) M_\text{CMF}$. Assuming an SFE $\epsilon = 0.3 \pm 0.1$ \citep{Alves2007} and a probability function for the distribution of masses, it would be possible to find the distribution of $n_*$. A similar study was performed by \citet{Holman2013} without a fixed value of SFE and with a mean value of $n_*$, $\bar{n}_*$. However, the degeneracy between SFE and $\bar{n}_*$ forces one of these parameters to be constrained. \citet{Hatchell2008} and \citet{Goodwin2008} used a \textit{\textup{fully multiple model}} that allows a fixed $\bar{n}_*$ in order to determine the SFE for cores. However, as clumps are much more massive and much more multiplicative than cores, it is not possible to make this link, and it is beyond the scope of this paper.

The environment, clustered and isolated, does not seem to change the slope of the CMF, the break masses do vary, with higher break masses found for clustered clumps. This indicates a difference in the histogram of $n_*$  or a variation in the SFE between clustered and isolated clumps.

A quick calculation provides a mean multiplicity factor ratio between clustered and isolated clumps, $\bar{n}_*,\text{R} = \bar{n}_*,\text{isolated}/\bar{n}_*,\text{clustered}$. By fixing the SFE to $\epsilon = 0.3$, $\bar{n}_*,\text{R}$ is defined as $\bar{n}_*,\text{R} = M_\text{CMF,isolated}/M_\text{CMF,clustered}$. The break masses, $M_1$ and $M_2$, are appropriate to compute $\bar{n}_*,\text{R}$. A range of $\bar{n}_*,\text{R}$ is found to be 0.4-0.6, meaning that $\bar{n}_*,\text{clustered} \approx 2 \times \bar{n}_*,\text{isolated}$.\\

\subsection{Comparison with environmental conditions}

\subsubsection{\ion{H}{II} regions}

One of the main stages of the high-mass star-forming process is the formation of \ion{H}{II} regions. In this section, the relationship between cluster candidates and \ion{H}{II} regions is investigated using the catalogues of \citet{Anderson2014} (And2014) and \citet{Paladini2003} (Pal2003), who detected a total of 8,400 and 1,142 \ion{H}{II} regions, respectively. Our
study region includes 7,589 of these \ion{H}{II} regions, with 6,561 and 1,008 from the And2014 and Pal2003 catalogues, respectively. \ion{H}{II} regions are only associated with cluster candidates if at least one of the clustered clumps is located within the radius of a \ion{H}{II} region. A total of 274 cluster candidates are associated with at least one \ion{H}{II} region, with 443 \ion{H}{II} regions associated with at least one cluster candidate. However, for a \ion{H}{II} region to be associated with the whole cluster candidate, it has to be as large as the cluster candidate. As a result of this, 178 cluster candidates are associated with 177 \ion{H}{II} regions. Of these \ion{H}{II} regions, 168 are from And2014, with the remaining 9 from the Pal2003 catalogue.\\
When we fold in the heliocentric distances from the And2014 catalogue, 54 cluster candidates are associated with 58 \ion{H}{II} regions. By considering large \ion{H}{II} regions, only 24 cluster candidates are associated with 19 \ion{H}{II} regions. However, by analysing the Herschel images, most of the clustered clumps seem to be associated with the photo-dissociation region (PDR) of the \ion{H}{II} regions independently of the use of the distance estimates. Most of the cluster candidates are larger than 20~pc, which implies an association with a large \ion{H}{II} region. As this case remains rare, it is an explanation why few associations can be found with large \ion{H}{II} regions.\\

\subsubsection{IRDCs}

The IRDCs are potentially associated with the early phases of high-mass star formation. Their association with cluster candidates is investigated by comparing to two IRDC catalogues. \citet{Simon2006} (Sim2006) used Midcourse Space Experiment data \citep{Mill1994, Egan1998} at 8.3\,$\mu$m, finding 10,931 IRDCs in the first and fourth quadrants, and \citet{Peretto2009} (Per2009) used the GLIMPSE and MIPSGAL surveys \citep{Benjamin2003,Carey2009} to obtain a catalogue of 11,303 IRDCs over the same Galactic longitude range. These catalogues are complementary, with a total of 17,109 unique IRDCs found over a combined catalogue as they overlap
by $\sim$ 20\% \citep{Peretto2009}.

Only 38\% of the clumps fall into the footprint of an IRDC, similar to the figure found by \citet{Billot2011}, but 333, or 67\%, of the cluster candidates are associated with IRDCs. As some IRDCs are hidden because of a high foreground luminosity, we can expect a better percentage of association.

\subsection{Comparison with other cluster catalogues}

The cluster candidates were compared with catalogues of clusters and star-forming regions from the literature. The literature clusters are mainly open stellar clusters, but some of them are embedded young stellar clusters. Table~\ref{refclustcat} lists the reference of the catalogues, the number of sources, and the number of sources in the cluster candidate catalogue range. We have a list of 704 stellar clusters \citep{Majaess2013, Morales2013, Solin2014} and 1,245 star-forming regions \citep{Avedisova2002, Lee2012, Beuther2002} in the range of the cluster candidates.

\begin{table}
\caption{Catalogues of clusters from the literature.$^{(a)}$Number of clusters in the whole catalogue.$^{(b)}$Number of clusters in the  range of the Hi-GAL product catalogue.} 
\label{refclustcat} 
\centering 
\begin{tabular}{c c c} 
\hline\hline 
Reference & $Nb_\text{clust}^{(a)}$ & $Nb_\text{clust}^{(b)}$  \\ 
 & & \\
\hline 
 \citet{Avedisova2002} & 3,235 & 1,049 \\ 
 \citet{Sridharan2002, Beuther2002} & 69  & 45 \\
 \citet{Lee2012} & 280  & 151 \\
 \citet{Majaess2013} & 230 & 41 \\
 \citet{Morales2013} & 695 &  531 \\
 \citet{Solin2014} & 160 &  132 \\
\hline
\end{tabular}
\end{table}

The matching radius between the stellar clusters and the cluster candidates was set at 3\arcmin and 9\arcmin of the central position of the cluster candidates. 3\arcmin is the peak size of the radii of the cluster candidates, estimated from the semi-major and semi-minor axes. Ninety-five percent of the cluster candidates have a radius smaller than 9\arcmin. For 3\arcmin, 50 literature stellar clusters are associated with 45 cluster candidates, and with a 9\arcmin search radius, 166 clusters are associated with 119 cluster candidates.\\
The low number of associations, 50 and 166 clusters compared to the total 704, may be due to the age of the clusters, some of which are open clusters, and will have migrated from the site at which they formed. This would make it difficult for them to associate with young clusters.\\

The matching radius between the star-forming regions and the cluster candidates was set at a greater value, of 15\arcmin, in addition to 9\arcmin. For 9\arcmin, 279 literature star-forming regions are associated with 199 cluster candidates, and with a 15\arcmin search radius, 697 literature star-forming regions, or 56\%, are associated with 325 cluster candidates. The association is difficult because some young and cold clusters trace new star-forming regions that were probably undetected until now. The large extinction towards the inner Galaxy can also prevent the detection of embedded stellar clusters, therefore the match between cluster candidates and star-forming regions or stellar clusters might be underestimated.

   %*****************************************************************************
   %********************                                     ********************
   %********************      Summary and perspectives       ********************
   %********************                                     ********************
   %*****************************************************************************

\section{Summary and perspectives}

The Hi-GAL physical properties catalogue has been analysed using the minimum spanning tree (MST) method to find over-densities on the sky, in 2D. This has occurred over Galactic latitudes of -70 to 67 deg. A total of 1,705 over-densities were found with at least ten members.\\

The heliocentric distance estimates (HDEs) were used to differentiate cluster candidates from potential cluster candidates. After recomputing, 1,633 over-densities was found, with 496 considered cluster candidates and 1,137 potential cluster candidates.\\
This is the largest catalogue of embedded clusters of nascent stars. This could help to identify star-forming regions for future studies, especially those with higher angular resolution.\\

The study was continued by analysing the spatial distribution of the cluster candidates, with almost all cluster candidates following the location of the spiral arms \citep{Englmaier2011}, and the latitude distribution was shown to follow the warp of the Galactic Plane. The spatial distribution of clumps within cluster candidates seems to follow a fractal distribution.\\

The CMF slopes of the isolated and clustered clumps are statistically indistinguishable from each other, which is consistent with the Kroupa IMF, implying that the IMF is set by the CMF and at the clump-formation stage. However, the break masses vary, which suggests a different SFE or mean multiplicity factor between these two environments.\\

We found that 55\% of the cluster candidates are associated with \ion{H}{II} regions and 68\% with IRDCs. A small number are associated with clusters from the literature.\\

Future work will explore the Outer Galaxy and also study individual sources and cluster candidates.

\begin{acknowledgements}

M Beuret thanks Arnaud Siebert for providing the shape of the spiral arms from \citet{Englmaier2011}.
This research has made use of the VizieR catalogue access tool, CDS, Strasbourg, France. The original description of the VizieR service was published in A\&AS 143, 23. This research has also made use of NASA's Astrophysics Data System Bibliographic Services. Davide Elia's and Eugenio Schisano's research activity is supported by the VIALACTEA project, a Collaborative project under Framework Programme 7 of the European Union funded under Contract \#607380, that is hereby acknowledged.

\end{acknowledgements}

%!!!!!!!!!!!!!!!!!!!!!!!!!!!!!!!!!!!!!!!!!!!!!!!!!!!!!!!!!!!!!!!!!!!!!!!!!!!!!!!!!!!!!!!!
%                                        BIBLIOGRAPHY
%!!!!!!!!!!!!!!!!!!!!!!!!!!!!!!!!!!!!!!!!!!!!!!!!!!!!!!!!!!!!!!!!!!!!!!!!!!!!!!!!!!!!!!!!

\bibliographystyle{aa} % style aa.bst
\bibliography{ref} % your references Yourfile.bib

%!!!!!!!!!!!!!!!!!!!!!!!!!!!!!!!!!!!!!!!!!!!!!!!!!!!!!!!!!!!!!!!!!!!!!!!!!!!!!!!!!!!!!!!!
%                                  APPENDIX
%!!!!!!!!!!!!!!!!!!!!!!!!!!!!!!!!!!!!!!!!!!!!!!!!!!!!!!!!!!!!!!!!!!!!!!!!!!!!!!!!!!!!!!!!

\begin{appendix}

\section{Distribution of mass}

\begin{figure*}
\centering
\resizebox{\hsize}{!}{\includegraphics{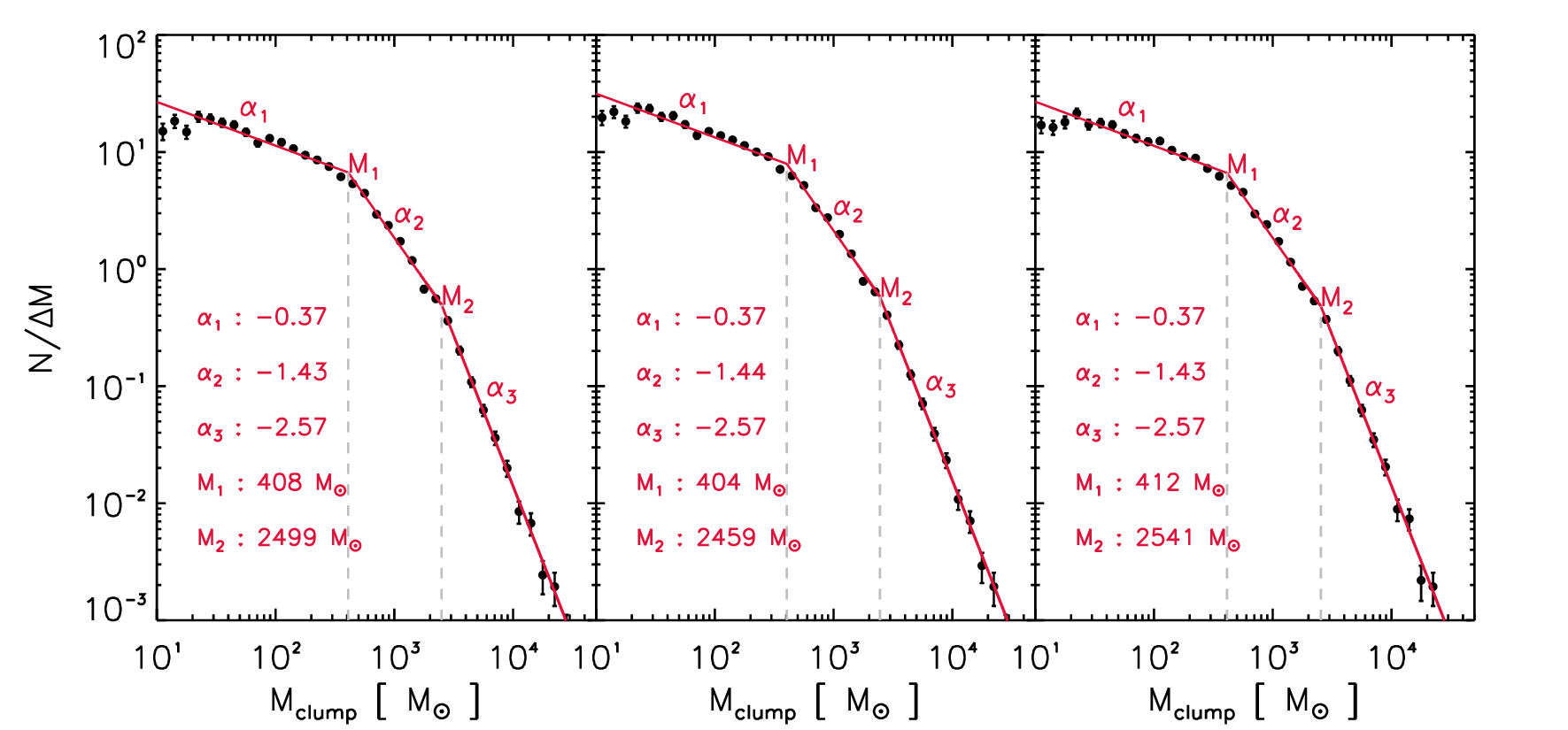}}
\caption{Mass distribution of clustered clumps with HDEs (left panel), with and without HDEs (middle panel), and without recomputing masses (right panel). For masses that are recomputed, we considered Eq.~\ref{equaM} and the ratio $D_\text{peak}^{2}/D^{2}$, where $D_\text{peak}$ is the distance of the cluster candidates and $D$ the individual distance of clumps. The two plots corresponds to the function $\Phi$ (Eq.~\ref{equaIMF}) versus mass. The red lines corresponds to the fitting segments on the CMF, which provide the slopes. $\alpha_1$, $\alpha_2$ , and $\alpha_3$ and the break masses, $M_1$ and $M_2$. The three considerations do not alter the shape or slopes of the CMFs.}
\label{masschangeclust}
\end{figure*}

\begin{figure*}
\centering
\resizebox{\hsize}{!}{\includegraphics{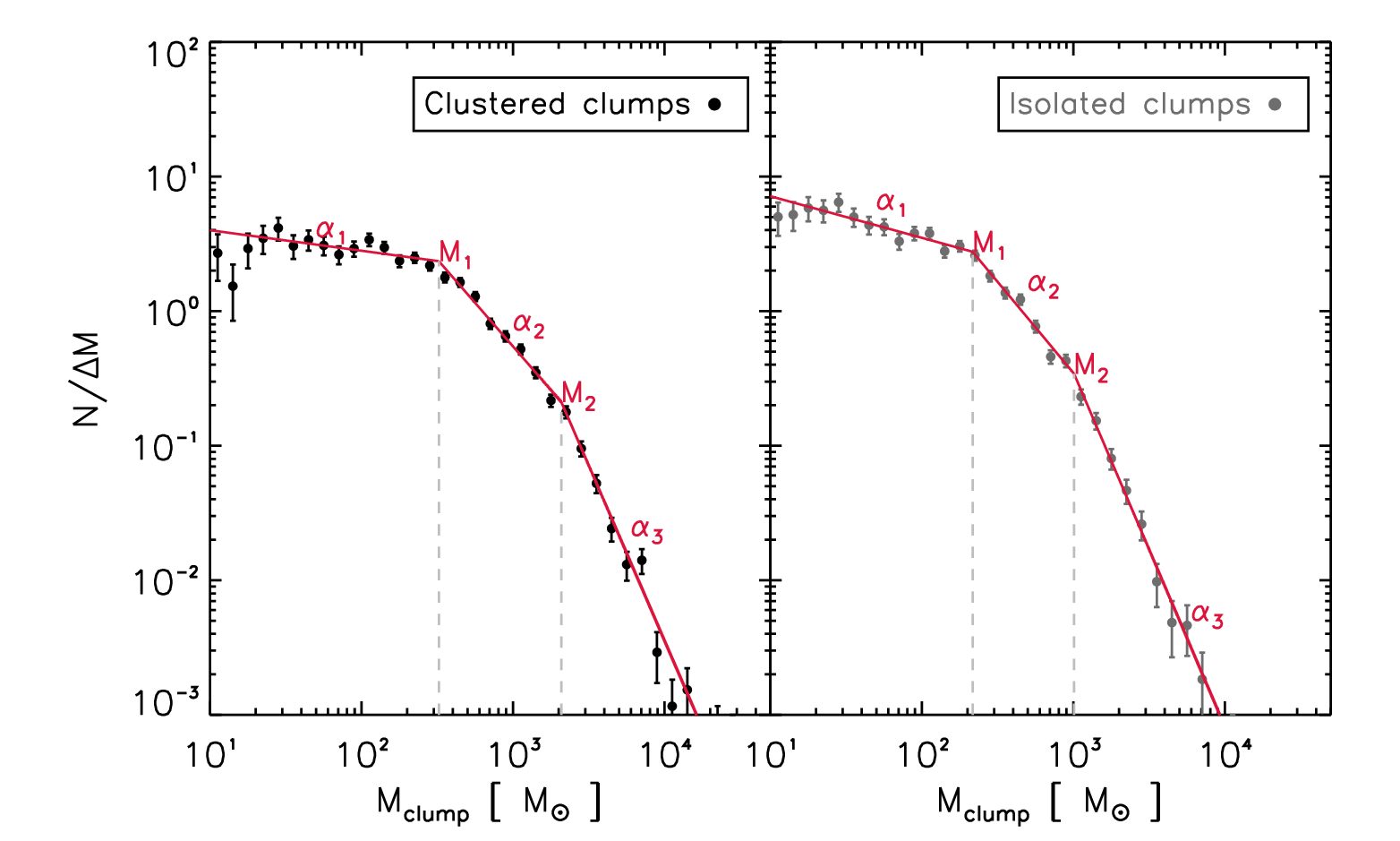}}
\caption{Mass distribution of clustered clumps (left panel) and isolated clumps (right panel) in the first quadrant. Only clumps with HDEs are used. The two plots correspond to the function $\Phi$ (Eq.~\ref{equaIMF}) versus mass. The red lines correspond to the fitting segments of the CMF.}
\label{isoclustfirstqd}
\end{figure*}

\begin{figure*}
\centering
\resizebox{\hsize}{!}{\includegraphics{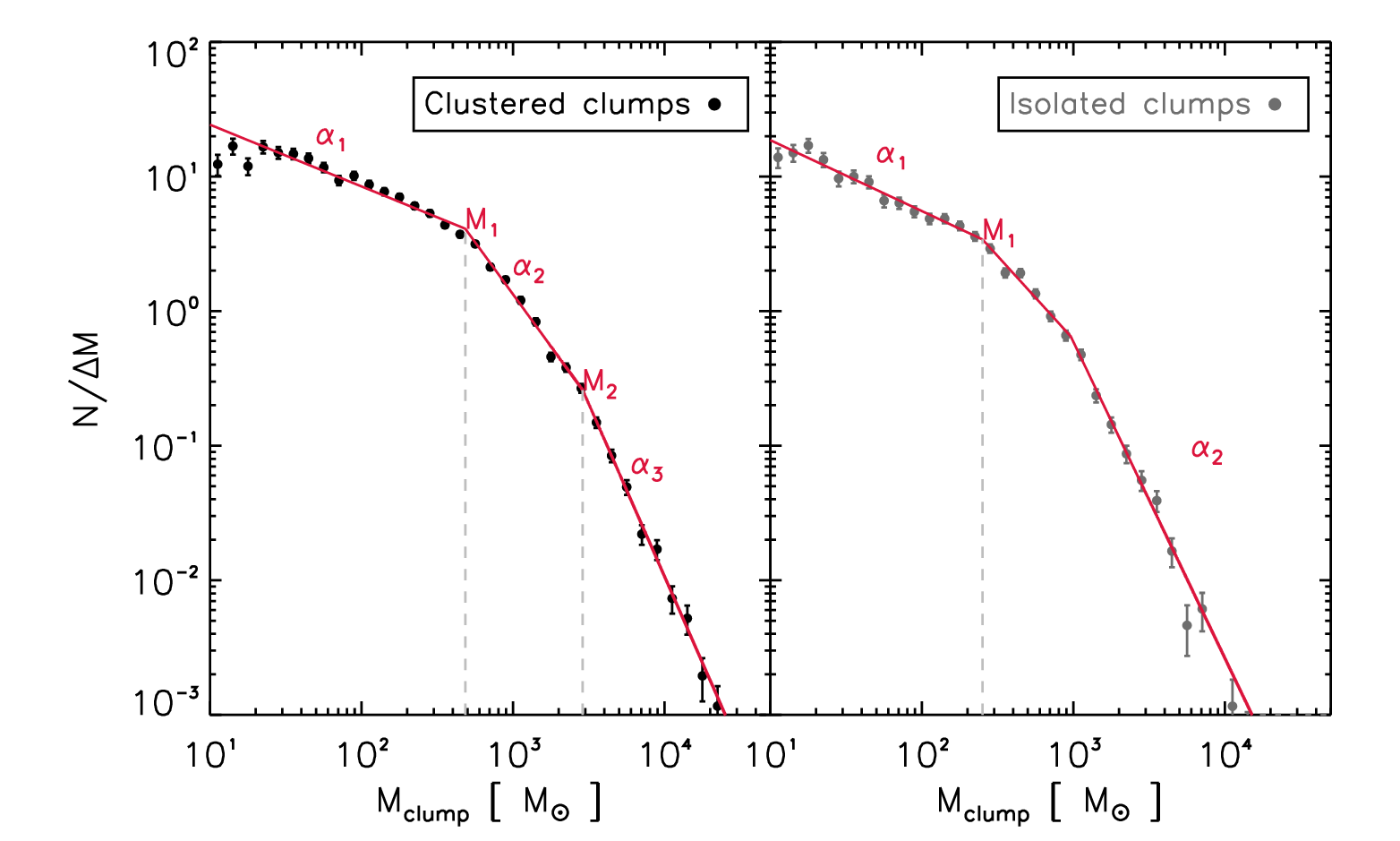}}
%\resizebox{\hsize}{!}{\includegraphics{figures/mass_preproto_clusteredonlydist_isolated_preproto_allhde_south_kroupa_2segment.ps}}
\caption{Mass distribution of clustered clumps (left panel) and isolated clumps (right panel) in the fourth quadrant. Only clumps with HDEs are used. The two plots correspond to the function $\Phi$ (Eq.~\ref{equaIMF}) versus mass. The red lines correspond to the fitting segments of the CMF.} %\textit{Bottom} : Same as the \textit{Top} but with a fit of 2 segments on the mass distribution of the isolated clumps.}
\label{isoclustfourthqd}
\end{figure*}

\end{appendix}

\end{document}